\documentclass[aps,pre,floatfix,nofootinbib,showpacs,twocolumn,amsmath,amssymb]{revtex4}
\usepackage{dcolumn}
\usepackage{bm}
\usepackage[dvips]{epsfig}
\usepackage{xcolor}
\usepackage{graphicx} \usepackage{amsmath} \usepackage{amssymb}
\graphicspath{ {images/} }
\newcommand{\comment}[1]{}
\newcommand{\BEQ}{\begin{equation}}
\newcommand{\EEQ}{\end{equation}}
\newcommand{\BEA}{\begin{eqnarray}}
\newcommand{\EEA}{\end{eqnarray}}

\renewcommand{\d}{{\rm d}}

\newcommand{\e}{\epsilon}

\newcommand{\om}{\omega}

\newcommand{\ba}{\bar{a}}

\newcommand{\bp}{\bar{p}}
\renewcommand{\ba}{\bar{a}}

\begin{document}

\title{Polymorphism in rapidly-changing cyclic environment \footnote{Published as A. E. Allahverdyan, S. G. Babajanyan, and C.-K. Hu
Phys. Rev. E {\bf 100}, 032401 (2019). }}

\author{Armen E. Allahverdyan$^{1)}$\footnote{armen.allahverdyan@gmail.com \\
$\dagger$ sanasar.babajanyan@gmail.com \\
$\ddagger$ huck@phys.sinica.edu.tw},
Sanasar G. Babajanyan$^{1)\,\dagger}$ and Chin-Kun Hu$^{2,3)\,\ddagger}$  }
\affiliation{$^{1)}$Yerevan Physics Institute,
Alikhanian Brothers Street 2, Yerevan 375036, Armenia\\
$^{2)}$ Institute of Physics, Academia Sinica, Nankang, Taipei
11529, Taiwan\\
$^{3)}$Department of Physics, National Dong Hwa University, Hualien 97401, Taiwan\\
}

\begin{abstract}

Selection in a time-periodic environment is modeled via the
continuous-time two-player replicator dynamics, which for symmetric
pay-offs reduces to the Fisher equation of mathematical genetics. For a
sufficiently rapid and cyclic [fine-grained] environment, the
time-averaged population frequencies are shown to obey a replicator
dynamics with a non-linear fitness that is induced by environmental
changes. The non-linear terms in the fitness emerge due to populations
tracking their time-dependent environment.  These terms can induce a
stable polymorphism, though they do not spoil the polymorphism that
exists already without them. In this sense polymorphic populations are
more robust with respect to their time-dependent environments. The
overall fitness of the problem is still given by its time-averaged
value, but the emergence of polymorphism during genetic selection can be
accompanied by decreasing mean fitness of the population. The impact of
the uncovered polymorphism scenario on the models of diversity is
examplified via the rock-paper-scissors dynamics, and also via the
prisoner's dilemma in a time-periodic environment. 

\end{abstract}







\maketitle

\section{Introduction}

Organisms live in a changing world and experience variations of biotic
and abiotic environmental factors. Hence environmental impact on
selection and evolution is an important research subject
\cite{levins,stein,bull,grant,kassen_minireview,wilke,dempster,janavar,svi,nagylaki_book,
nagylaki,kimura,li,gill,clark,strobeck,miner_vonesh,winn,jasmin,cook}
Populations respond to an inhomogeneous environment by developing
polymorphism, where two or more different morphs
exist in one interbreeding population
\cite{levins,stein,bull,grant,kassen_minireview}. 

\comment{ Besides environmental heterogeneity there are other sources of
polymorphism, such as frequency-dependent selection, heterozygote
advantage and mutations.  Polymorphism can be restricted to the
phenotype level, or it may be controlled genetically, e.g., by multiple
alleles at a single locus; e.g.  human ABO blood groups \cite{grant}. }

Here are two known examples of polymorphism related to a time-periodic
environment \cite{grant,cook}.
Populations of the land snail ({\it Cepaea Nemoralis}) consist of three
morphs having respectively brown, pink and yellow coloured shells.  The
shell colour is regulated by multiple alleles of one gene; the brown
(yellow) allele is the most dominant (recessive) one.  For a population
living in a forrest, the brown and pink morphs have a selective
advantage at the spring time, since they correlate with the colour of
the ground laying making the snail less visible for predators
\cite{grant,cook}. The yellow morph has an advantage at summer and
autumn on the yellow-green laying.  In addition, the yellow morph is
more resistant to high and low temperatures \cite{grant,cook}.  Another
example of polymorphism is the two basic morphs of the spotted lady
beetle ({\it Adalia bipunctata}), which carry black (with red spots) and
red (with black spots) colour, respectively \cite{zakharov,lady}.  This
polymorphism relates to the fact that the black morph is more resistant
to extreme conditions (at winter and summer, or to industrial stresses),
while the red morph does better under normal conditions
\cite{zakharov,lady}. 

In a slow (coarse-grained) environment each individual 
sees mainly one fixed environment, which can change from one
generation to another \cite{levins}.  A rapidly-changing (i.e fine
grained) environment changes many times during the life-time of each
individual; see the above example of {\it Cepaea} and note that this
snail lives seven to eight years \cite{grant,cook}. A population moving
via spatially-inhomogeneous environment faces the same problems as a
static population in a time-dependent environment. 

Much attention was devoted to modeling polymorphism in slow
environments
\cite{levins,stein,bull,kassen_minireview,dempster,janavar,svi,nagylaki,gill,clark}.
The main outcome is that morphs in a slow environment are governed by
the geometric mean of their fitness \footnote{The difference between the
geometric and arithmetic average can be illustrated as follows.  The
discrete-time logistic growth $N_{k+1}=\phi_k N_k $ leads to
$N_k=e^{\sum_{l=1}^K \ln \phi_l}N_0$; thus $\phi_k$ contributes into the
long-time fitness multiplicatively. Hence for $K\gg 1$ we get:
$\frac{1}{K}\ln[N_k/N_0]=\langle \ln \phi\rangle$, i.e. the geometric
fitness average.  In contrast, the continuous-time logistic growth
$\dot{N}=f(t)N$ integrates as $N(t)=e^{\int_0^t \d s f(s)}N(0)$, and
$f(s)$ contributes to the long-time fitness additively. Hence we get the
arithmetic average for $T\gg 1$: $\frac{1}{T}\ln[N(T)/N(0)]=\langle
f\rangle$.  } \cite{aversion}. 

\comment{Polyphenism is an irreversible (within a single individual)
phenotypic polymorphism which does have a clear seasonal periodicity,
e.g., the sex determination in alligators is triggered by environmental
factors \cite{ford}. }

\comment{
Two scenarios were proposed in fine-grained environments: either
the evolution selects a generalist, which adapts to the average
environment and thus does satisfactorily in all its patches, or|due to
constraints on adaptation to the average environment|a specialist
emerges, which has adapted to one environmental patch, but does badly in
other patches \cite{levins,kassen_minireview}. }

Rapidly-changing (i.e. fine-grained) environments got less attention so
far. Here the effective fitness is postulated to be given by arithmetic
time-averages of selection coefficients
\cite{levins,strobeck,stein,kassen_minireview,wilke}. This is
incomplete, because correlations between the fitness and the state of
population are not accounted for. Experiments on the evolution in a
rapidly-changing environment show that evolving populations can respond
to time-varying aspects of their environment
\cite{miner_vonesh,winn,jasmin}. The total fitness during such processes
need not increase \cite{jasmin}.  It is also known that individual
organisms can develop different phenotypes in response to different
environmental conditions \cite{ford}. In particular, the phenotypic
difference can be reversible within a single individual, and can have a
well-defined seasonal periodicity \cite{drent}.  Thus a theoretical
model is needed to explain how rapidly-changing environment leads to
polymorphism, and how it relates to the total fitness of the population
\cite{prl}. 

Here we present a theory for polymorphism in rapidly-changing,
time-periodic environment based on (continuous-time) Evolutionary Game
Theory (EGT) that describes competition of several morphs
\cite{hofbauer,zeeman}.  This approach unites genetic and phenotypic selection
models into a single and flexible formalism. In a particular case, EGT
leads to continuous-time equations of classical mathematical genetics
\cite{hofbauer}. We show that in addition to the contribution given by
time-averaged selection coefficients, there are also non-linear (and
non-perturbative) terms in the effective fitness that emerge due to
tracking by competing morphs of their time-dependent environment. These
non-linear terms predict new regimes of polymorphism.  A polymorpshim
that already exists on the level of the time-averaged selection
coeffients is not eliminated. For genetic selection the overall
fitness|which is shown to be equal to the time-averaged fitness|does not
need to increase in time during the establishment of the polymorphism.
Our results agree with necessary conditions for new
regimes in environmentally driven population dynamic equations
\cite{levins79,fox}.

This paper is organized as follows. Section \ref{rgt} recalls the
replicator dynamics. Section \ref{ee} deduces our main result: a
rapidly-changing environment brings in non-linear (multi-party) fitness.
These non-linear terms can be larger than those given by the average
linear fitness. Section \ref{fito} discusses the time-averaged and
effective fitness. In sections \ref{two} and \ref{three} we present our
results on the polymorphism for the two- and three-morph situation,
respectively.  Section \ref{two} also connects with genetic selection
models in a rapidly-changing (fine-grained) environment. Section
\ref{remind} studies the famous prisoner's game in a time-periodic
environment and shows in which concrete sense such an environment leads
to resolving the prisoner's dilemma.  Section \ref{three} discusses the
environment-induced polymorphism in the rock-paper-scissors dynamics,
which is one of the main models for studying biodiversity in interacting
populations.  Section \ref{Poincare} outlines general aspects of the
polymorphism emergence using the notion of Poincar\'e indices. Section
\ref{summa} summarizes our results and outlines open problems.

\section{Replicator dynamics}
\label{rgt}

Evolutionary Game Theory
(EGT) describes interacting agents separated into several groups (morphs)
\cite{svi,hofbauer,zeeman,namara}. The reproduction of each group is governed by
its fitness, which depends on interactions between the morphs. The
replicator dynamics approach to EGT describes the
time-dependent frequency $p_k(t)$ of the group $k$, which is the
number of agents $N_k$ in the group $k$, over the total number of
agents in all $n$ groups: $p_k ={N_k}/{\sum_{k=1}^n N_k}$. The (Malthusian)
fitness $f_k$ of the group $k$ is a linear function of the
frequencies \cite{svi,hofbauer,zeeman}:
\BEA
\label{z1}
f_k(a,p)={\sum}_{l=1}^n
a_{kl} p_l, \qquad k=1,\ldots, n,
\EEA
where the {\it payoffs}
$a_{kl}$ (selection coefficients) account for the interaction between (the agents from)
groups $k$ and $l$. The replicator dynamics \cite{hofbauer,zeeman}
facilitates the (relative) growth of groups with fitness larger
than the mean fitness ${\sum}_{l=1}^n p_l f_l$:
\BEA
\label{z2}
\dot{p}_k=p_k [\,f_k(a,p)-{\sum}_{l=1}^n p_l f_l\,]\equiv
G_k[a,p].
\EEA
Note that the same fitness $f_k(a,p)$ can be introduced for the number
of agents in each group: $\dot{N}_k=N_k f_k (p)$. This is the equation
for the logistic growth, but with frequency-dependent growth rates.  We
revert to (\ref{z2}) after substituting $p_k=N_k/(\sum_{l=1}^n N_l)$.
Now polymorphism means a
stable state, where two or more $p_k$ are non-zero.

There are several applications of replicator dynamics:

--
Animal (agent contest), where the groups correspond to the
strategies of agent's behavior, while $p_k$ is the probability by which
an agent applies the strategy $k$ \cite{hofbauer,zeeman}.  Alternatively, there
are different agents each one applying one fixed strategy. Agents
applying similar strategies can be joined into groups, and then
(\ref{z2}) describes the evolution of the relative size of those groups
\footnote{These two situations|different groups of agents, each agent
applying a fixed strategy, or a single agent applying various
strategies|are not always equivalent \cite{thomas}.}. The actual
mechanism by which $p_k$ changes depends on the concrete implementation
of the model (inheritance, learning, imitation, infection, {\it etc}). 

--
Selection of genes, where $p_k$ is the frequency of one-locus
allele $k$ in panmictic, asexual, diploid population, and where
$a_{kl}=a_{lk}$ refers to the selective value of the phenotype driven by
the zygote $(kl)$ \cite{svi}.  Then (\ref{z1}, \ref{z2}) are the Fisher
equations for the selection with overlapping generations
\cite{svi,hofbauer}. Eq.~(\ref{z2}) applies to lower-level organisms
(such as bacteria), which reproduce almost continuously with each
generation bringing a small contribution to the overall population.  The
situation with (for instance) a human population, where there are no
breeding seasons (reproduction is continuous) and there are strong
overlaps between generations also roughly corresponds to 
(\ref{z2}). For a more precise description of
this situation one should employ generalizations of (\ref{z2}), which
explicitly account for the age of each generation; see
\cite{nagylaki_book,svi} for details. Whenever generations overlap, but
there is a definite breeding season, one still expects the applicability
of (\ref{z2}) \cite{nagylaki_book}. 

--
Within a two-party game theory between players $I$ and $II$, strategies
refers to $k=1,\ldots, n,$ in (\ref{z1}). The pay-off $a_{kl}$ is
what $I$ gets, if $I$ and $II$ act $k$ and $l$, respectively.  The
replicator approach (\ref{z1}) applies to symmetric games
\cite{hofbauer,zeeman}, where $a_{kl}$ is also what $II$ gets, if $I$ and $II$
act $k$ and $l$, respectively. In such symmetric games, both $I$ and $II$
apply strategy $k$ with the same probability $p_k$ \cite{hofbauer,zeeman}.


\section{Effective equations for rapidly-changing environment}
\label{ee}

\subsection{Derivation}
\label{deriv}

We consider a varying, but predictable environment, which acts on the
phenotypes making $a_{kl}$ periodic functions of time with a {\it
single} period $2\pi/\om$ \cite{svi}:
\BEA
\label{corrida}
a_{kl}(\tau )=a_{kl}(\tau+2\pi), \qquad \tau\equiv\om t.
\EEA
There are well-defined
methods to decide to which extent a varying environment is
predictable for a given organism \cite{clark}. The oscillating
payoffs can reflect the fact that different morphs (alleles,
strategies) are dominating at different times.

We assume that the environment is fast [fine-grained]: the time-averaged
(systematic) change of the population structure over the environment
period ${2\pi}/{\om}$ is small. We separate the time-dependent payoffs
$a_{kl}(\tau )$ into the constant part $\bar{a}_{kl}$ and the
oscillating part $\widetilde{a}_{kl}(\tau)$:
\BEA \label{6}
a_{kl}(\tau)=\bar{a}_{kl}+\widetilde{a}_{kl}(\tau).
\EEA
The time-average of the oscillating part is zero:
\BEA
\label{66}
\overline{\widetilde{a}_{kl}}\equiv \int_0^{2\pi}\frac{\d
\tau}{2\pi}\, \widetilde{a}_{kl}(\tau)=0.
\EEA
Formulas similar to (\ref{66}) will define below the time-averages of other periodic functions of $\tau$.

For future purposes let also $\hat{a}_{kl}$ be the primitive of
$\widetilde{a}_{kl}$ with its time-average equal to zero:
\BEA
\partial_\tau\hat{a}_{kl}(\tau)=\widetilde{a}_{kl}(\tau), \qquad \overline{\hat{a}_{kl}}=0.
\label{6.1}
\EEA
Now $\overline{\hat{a}_{kl}}=0$ makes $\hat{a}_{kl}(\tau)$ 
periodic with the same period $2\pi$; cf.~(\ref{corrida}). Below we shall frequently use:
\BEA
\overline{\, \widetilde{a}_{kl} \,\hat{a}_{kl}\,}=0.
\label{darder}
\EEA

Following the Kapitza method \cite{ll,kapitza}, we represent $p_k$ as a
slowly varying part $\bp_k$ plus $\e_k[\bp (t),\tau]$. The latter is
smaller than $\bp_k$, oscillates fastly on the environment time
$\tau$, and averages to zero:
\BEA \label{8}
&&p_k(t)=\bp_k(t)+\e_k[\bp (t), \tau],\\
\label{8.1}
&&\bar{\e}_k[\bp (t)]=\int_0^{2\pi}\frac{\d
\tau}{2\pi}\, \e_{k}[\bp (t),\tau]=0,
\EEA
where the time-average is
taken over the fast time $\tau$ for a fixed slow time $t$. Note
that the fast $\e_k$ depends on the slow $\bp$.

Now put (\ref{8})
into (\ref{z2}) and expand the right-hand-side of (\ref{z2}) over $\{\e_k\}$:
\begin{gather}
\dot{\bp}_k+\dot{\bp}_\alpha\partial_\alpha \e_k+\om \partial_\tau
\e_k =[1+\e_\alpha\partial_\alpha +{\cal
O}(\e^2)]G_k[a(\tau),\bp],
\label{z5}
\end{gather}
where the summation over the repeated Greek indices is assumed,
\BEA
\label{olu}
A_\alpha B_\alpha \equiv {\sum}_{\alpha=1}^n A_\alpha B_\alpha,
\EEA
and where
\BEA
\label{kolu}
\partial_\alpha X\equiv \frac{\partial }{\partial \bp_\alpha}X.
\EEA

The fast factor $\e_i$ is searched for via expanding over
$\frac{1}{\omega}$:
\BEA
\e_k= \frac{1}{\omega}
\e_{k,1}+\frac{1}{\omega^2} \e_{k,2}+\ldots.
\EEA
Substitute this into (\ref{z5}) and recall that $G_k$ is linear over $\{a_{kl}\}$:
\BEA
\dot{\bp}_k&+& \frac{1}{\omega}\dot{\bp}_\alpha\partial_\alpha \e_{k,1}+\partial_\tau
\e_{k,1}+\frac{1}{\omega}\partial_\tau \e_{k,2} =G_k[\bar{a}(\tau),\bp] \nonumber\\
&+&G_k[\widetilde{a}(\tau),\bp] +\frac{1}{\omega}\e_{\alpha,1}\,\partial_\alpha \,G_k[a(\tau),\bp]+{\cal O}\left(
\frac{1}{\omega^2}\right). ~~~~~
\label{condor}
\EEA
Eq.~(\ref{condor}) contains terms with different time-scales and of different orders of magnitude. First we
select terms which are of order ${\cal O}(1)$ and vary on the fast time-scale:
\BEA
\label{z8z}
\partial_\tau
\e_{k,1}=G_k[\widetilde{a}(\tau),\bp(t)]+{\cal O}(\frac{1}{\omega}).
\EEA
This equation can be integrated straightforwardly, because
the characteristic time $\tau$ and the slowly-changing variable $\bp(t)$ are separated from each other
[see (\ref{8})]:
\BEA
\label{z8}
\e_k[\bp(t), \tau]&=&\frac{1}{\om}\e_{k,1}+{\cal O}(\frac{1}{\omega^2})\nonumber\\
&=& \frac{1}{\om}\, G_k[\hat{a}(\tau),\bar{p}(t)]
+{\cal O}(\frac{1}{\omega^2}),
\EEA
where $\{\hat{a}_{kl}(\tau)\}$ is defined in (\ref{6.1}). Note from (\ref{z8}) that the normalization $\sum_k \e_{k,1}=0$
demanded by (\ref{8}) does hold.

Once (\ref{z8}) is separated out, we average the remainder in (\ref{z5}) over the fast time $\tau$ and obtain the
evolution of slow terms
\begin{gather}
\label{9}
\dot{\bp}_k=G_k[\bar{a}, \bp]
+\overline{\, \e_\alpha[\bp,\tau] \,\,\partial_\alpha
G_k[\widetilde{a}(\tau), \bp]  \, }+{\cal O}(\frac{1}{\omega^2})\\
=G_k[\bar{a}, \bp]
+\frac{1}{\omega}\overline{\,G_\alpha[\hat{a}(\tau),\bar{p}(t)] \,\partial_\alpha
G_k[\widetilde{a}(\tau), \bp]  \, }+{\cal O}(\frac{1}{\omega^2}),
\label{zaza}
\end{gather}
where the time-average is defined as in (\ref{8.1}).

Eq.~(\ref{zaza}) is a closed equation for the averaged (slowly-changing) quantities $\bar{p}_k$.
When working out $\overline{\, \e_\alpha[\bp,\tau] \,\,\partial_\alpha
G_k[\widetilde{a}(\tau), \bp]  \, }$ in
(\ref{9}) we employ $\overline{\hat{a}\widetilde{b}}=-\overline{\widetilde{a}\hat{b}}$,
note the following simplifying points:
\BEA
\sum_{nl}\overline{\hat{a}_{kn}\widetilde{a}_{kl} }\,\bp_n\bp_l=
\sum_{klmn}\overline{\hat{a}_{kl}\widetilde{a}_{mn} }\,\bp_k\bp_l\bp_m\bp_n=0,
\EEA
and get again a replicator equation
\BEA
\label{11}
\dot{\bp}_k
&=&\bp_k\left[{\cal F}_k(\bp)-\bp_\alpha {\cal F}_\alpha(\bp)\right],\quad k=1,...,n, \\
\label{12}
{\cal F}_k(\bp) &=&
\bar{a}_{k\alpha} \bp_\alpha
+b_{k\alpha\beta} \bp_\alpha\bp_\beta
+c_{k\alpha\beta\gamma} \bp_\alpha\bp_\beta\bp_\gamma,
\EEA
where ${\cal F}_k$ is the effective (already non-linear) fitness, and
\BEA
\label{14}
b_{klm}\equiv\frac{1}{\om}\,
\overline{\widetilde{a}_{kl}\,\hat{a}_{lm}},
\quad
c_{klmn}\equiv \frac{1}{2\om}\,
\overline{\,\hat{a}_{kl} \,\, [\widetilde{a}_{mn}+\widetilde{a}_{nm}] \, },
\EEA
account for (new) non-linear terms in ${\cal F}_k$. 
Eqs.~(\ref{11}, \ref{12}) is our central result.

\comment{(a ready part)
Note an obvious generalization of the above derivation.
Instead of (\ref{z2}) we can start from a more general equation
\BEA
\label{muto}
\dot{p}_k=p_k [\,f_k(a,p)-{\sum}_{l=1}^n p_l f_l\,]
+\sum_l(\mu_{kl}p_l-\mu_{lk}p_k),
\EEA
where $\mu_{kl}$ are conditional probability rates of mutations.
If we make a plausible assumption that $\{\mu_{kl}\}$ do not depend
on environmental variations, then
the term $\sum_l(\mu_{kl}\bp_l-\mu_{lk}\bp_k)$ is to be added
to the right-hand-sides of (\ref{9}, \ref{zaza}).
}

Expectedly, the fast environment contributes the time-averaged payoffs
$\bar{a}_{kl}$ into ${\cal F}_k$ \cite{levins}. In addition, each group
$k$ gets engaged into three- and four-party interactions with payoffs
$b_{klm}$ and $c_{klmn}$, respectively. Recalling our discussion after
(\ref{z2}), we can interpret $\sum_{lm}b_{klm}\bp_l\bp_m$ in (\ref{12})
as the average pay-off received by one of three players upon applying
strategy $k$. 

The terms with $b_{klm}$ and $c_{klmn}$ in (\ref{12}) exist due to
tracking of the environment by the morphs; see (\ref{8}, \ref{zaza}).
These terms need not be small as compared to $\bar{a}_{kl}$-terms, since
the derivation of (\ref{11}--\ref{14}) applies for $\bp_k\gg \e_k$,
which can hold even for $\bar{a}_{kl}\to 0$. The next-order (omitted)
terms in (\ref{12}) are already perturbative, i.e. they have to be
smaller than the terms that were kept in (\ref{12}). 

We get $b_{klm}=c_{klmn}=0$ [due to (\ref{darder})], if only one $a_{kl}$
varies in time, or if all $\widetilde{a}_{kl}$ oscillate at one phase:
\BEA
\label{mashad}
\widetilde{a}_{kl}=a(\omega t)\xi_{kl}, 
\EEA
where $\xi_{kl}$ are constant amplitudes. I.e. the non-linear terms
[with $\propto b_{klm}$ and $\propto c_{klmn}$] are non-zero due to
interference between the environmental oscillations of
$\widetilde{a}_{kl}$ and those of $\e_k$, which are delayed over the
environmental oscillations by phase $\pi/2$; see (\ref{z8}). 

For the existence of non-linear terms we also need the
frequency-dependent selection; e.g.  no non-linear terms similar to
(\ref{14}) will be present for non-interacting replicators
\BEA
\label{ded}
\dot{p}_k=p_k[a_k(t)-a_\alpha(t) p_\alpha],
\EEA
because the fitness $p_k a_k(t)$ is a linear function of $p_k$. For this
example $\bp_k$ are determined only by the averages $\bar{a}_k$.  Both
the emergence of the non-linear terms and their absence for (\ref{ded})
agree with general necessary conditions found in \cite{levins79} for
potentially new regimes in environmentally driven population dynamic
equations; see \cite{fox} for a recent review. 

Now $b_{klm}$ and $c_{klmn}$ that can potentially lead to new mechanisms
of polymorphism, disappear for a very fast environment
$\omega\to\infty$; see (\ref{14}). This roughly agrees with the
Intermediate Disturbance Hypothesis, which is well-known in ecology and
population dynamics and which states that the diversity in coexisting
populations is facilitated by intermediate (for our situation not very
fast) enviromental changes \cite{fox} \footnote{Ref.~\cite{fox} by Fox
critically discusses the empiric support of the intermediate disturbance
hypothesis, and opines that such a support is mostly lacking.  In
response, Sheil and Burslem \cite{sheil} argued that the empirical
support is there, but the hypothesis has to be formulated correctly. We
note that \cite{fox} contains a lucid discussion on various theoretical
mechanisms by which inhomogeneous environment can lead to diversity.
But the theoretical review in Ref.~\cite{fox} does not focus
on environmental time-scales. Our results broadly agrees with
restrictions reviewed in Ref.~\cite{fox}. }.

Finally, we note that whenever the non-linear fitness terms disappear,
e.g. due to (\ref{mashad}), one can try to go to the order of ${\cal
O}(\frac{1}{\omega^2})$ terms. Appendix \ref{mansur} shows that this way
does not lead to a theory that is useful for polymrorphism scenarios. 
One reason for this is that|in contrast to non-linear terms
$b_{klm}$ and $c_{klmn}$ in (\ref{11}--\ref{14})|the ${\cal
O}(\frac{1}{\omega^2})$-terms have to be smaller than the average
fitness terms $\bar{a}_{k\alpha} \bp_\alpha$.

\subsection{Initial conditions}
\label{inici}

\comment{(ready part)
Recall that (\ref{8}) means separation of the full dynamics of $p_k(t)$
into two parts: slow variable $\bp_k(t)$ and fastly oscillating part
$\e_k$. For the slow variable we derived equations (\ref{11}, \ref{12}).
However, provided that we are given initial conditions $p_k(0)$, we
still have to indicate with which initial conditions equations
(\ref{11}, \ref{12}) are to be used. This is an important question,
because we shall see below that different initial conditions for $\bp_k$
(and thus for $p_k$) will converge with time to different attractors.
Once the attraction basins will be determined from (\ref{11}, \ref{12})
in terms of $\bp_k(0)$, we need to know them in terms of original initial
conditions $p_k(0)$.}

Looking at (\ref{8}, \ref{z8}) one notes that
\BEA
\label{poncho}
p_k(0) = \bp_k (0) +\frac{1}{\om}\, G_k[\hat{a}(0),\bar{p}(0)]+{\cal O}(\frac{1}{\omega^2}),
\EEA
which expresses $p_k(0)$ in terms of $\bp_k(0)$. The meaning of (\ref{poncho}) is as follows.

When the dynamics is switched on at the initial time $t=0$, $p_k(t)$
converges within few oscillation periods from $p_k(0)$ to a function
with average $\bp_k(0)$. Once this converges is over, $\bp_k(t)$ changes
slowly.  The difference between $p_k(0)$ and $\bp_k(0)$ is seen from
(\ref{poncho}) to depend on the initial phases of the oscillating
functions $a_{ln}(t)$, contrary to an intuitive expectation that these
phases will completely irrelevant for the long-time dynamics. They will
be indeed irrelevant provided that the system posses only one (stable)
rest point.  Otherwise, if there are several attraction basins, the
difference between $p_k(0)$ and $\bp_k(0)$ will matter at least for those initial
conditions which are close to the boundary between two basins; see below for more
illustrations. This difference is
known as the ``initial slip". It was studied for several classes of
dynamical systems possesing time-scale separation; see
\cite{davidson,slips}.

\subsection{Local stability of vertices} \label{barra} The vertex
points, where all $\bp_k$'s besides one equal zero, are rest points of the
effective replicator equation (\ref{11}--\ref{12}). Moreover, the
non-linear terms $b_{klm}$ and $c_{klmn}$ do not have influence on the local
stability of a vertex, in the sense that they do not alter the
eigenvalues of the Jacobian at a vertex. To see this, assume that the
vertex is given by $\bp_1=1$, and also assume that the independent
variables over which the Jacobian is to be taken are
$\bp_1,\ldots,\bp_{n-1}$. When the Jacobian matrix
\BEA
\label{kho}
\left\{{\cal J}_{kl}\right\}_{k,l=1}^{n-1},\qquad
{\cal J}_{kl}=\frac{\partial }{\partial \bp_l}\left(\bp_k\left[{\cal F}_k(\bp)-\bp_\alpha {\cal F}_\alpha(\bp)\right]\right),
\EEA
is taken at the vertex, we get
\begin{align}
\left.  {\cal J}_{kl}\right|_{\bp_1=1} =& \delta_{kl} \left. [\,{\cal F}_k(\bp)-{\cal F}_1(\bp)\,] \right|_{\bp_1=1}\nonumber \\
+& \delta_{k1} \left. [\,{\cal F}_n(\bp)-{\cal F}_l(\bp)\,] \right|_{\bp_1=1}.
\end{align}
Now note from (\ref{14}):
\BEA
\left. {\cal F}_{l}\right|_{\bp_1=1}=\bar{a}_{l 1}
+b_{l 11} +c_{l 111}=\bar{a}_{l 1}, \quad  l =1,\ldots,n.
\label{buf}
\EEA
Eqs.~(\ref{kho}--\ref{buf}) show that the Jacobian matrix (and hence its
eigenvalues) at the vertex do not depend on the non-linear terms
$b_{klm}$ and $c_{klmn}$.  We stress that this conclusion holds due to
the specific form (\ref{14}) of the non-linear terms, i.e. the
conclusion need not hold if new terms in fitness are introduced for some
{\it ad hoc} reason. It will play a role in understanding general
implications of these terms; see section \ref{Poincare}. 

\section{Fitness: time-average versus effective}
\label{fito}

Equation~(\ref{11}) implies that the relative growth of two morphs
is determined by the effective fitness difference:
\BEA
\frac{\d}{\d t}(\bp_k/\bp_l)=(\bp_k/\bp_l)({\cal
F}_k-{\cal F}_l).
\EEA
Thus in {\it stable} rest points of (\ref{11})
the (effective) fitness of surviving morphs are equal to each
other, while the fitness of non-surviving ($\bar{p}_k\to 0$)
morphs is smaller (Nash equilibrium) \cite{hofbauer,zeeman}.

Another pertinent quantity is the time-averaged fitness
$\overline{f_k[a(t), p(t)]}$; see (\ref{z1}, \ref{66}). Employing
(\ref{8}, \ref{8.1}, \ref{z8}) we deduce that this quantity is equal to the effective fitness
modulo small corrections of order of ${\cal O}(\omega^{-2})$:
\BEA
{\cal
F}_k=\overline{f_k(\bar{a}+\widetilde{a}, \bp+\e)}+{\cal O}(\omega^{-2}).
\EEA
The overall fitness of the population is characterized by the mean
effective fitness $\bp_\alpha {\cal F}_\alpha=\sum_{\alpha=1}^n
\bp_\alpha {\cal F}_\alpha$, which is also equal to its time-averaged analogue:
\begin{align}
&\Phi
\equiv\overline{\,\,(\bp_\alpha+\e_\alpha)(\bar{a}_{\alpha\beta}+\widetilde{a}_{\alpha\beta})
(\bp_\beta+\e_\beta)\,\,}=\bp_\alpha {\cal F}_\alpha+{\cal O}(\frac{1}{\omega^{2}}), \nonumber\\
& \bp_\alpha {\cal F}_\alpha = \bar{a}_{\alpha\beta}\bp_\alpha\bp_\beta
+b_{\alpha\beta\gamma}\bp_\alpha\bp_\beta\bp_\gamma.
\label{kondor}
\end{align}
The contribution $c_{\alpha\beta\gamma\delta}\bp_\alpha\bp_\beta\bp_\gamma\bp_\delta$
in $\bp_\alpha {\cal F}_\alpha$ nullifies due to (\ref{14}) and
$\overline{\hat{a}\widetilde{b}}=-\overline{\widetilde{a}\hat{b}}$.

The mean fitness $\Phi$ is especially important for genetic selection,
$a_{ik}(t)=a_{ki}(t)$, since for the constant payoff situation
($a_{ik}=a_{ki}$ do not depend on time) in the replicator equation
(\ref{z2}), $\Phi$ monotonically increases towards its nearest local
maximum over the set of variables $\bp_k$
\cite{svi,nagylaki_book,hofbauer}. This is the fundamental theorem of
natural selection. 

It may seem that (\ref{kondor}) recovers the known statement by Levins
that the {\it effective} overall fitness in a rapidly-changing (fine-grained) environment
is given by the average fitness \cite{levins} \footnote{Levins derived
this statement from several qualitative assumptions. As noted by
Strobeck, this statement is basically a postulate, and need not hold for
all reasonable models of rapidly-changing environments \cite{strobeck}. }.
The essential point made by Levins that the average fitness increases in
time (for rapidly-changing environments and symmetric pay-offs). However, we
shall see below that due to non-linear terms (\ref{14}) the effective (=
averaged) fitness $\Phi$ can decrease in time, specifically when the
environment-induced polymorphism is essential.  Thus, the fundamental
theorem of natural selection can be violated in rapidly-changing
environment, despite of the fact that the effective fitness is equal to the
time-averaged fitness. 


\begin{figure}
\vspace{0.3cm}
\begin{center}
\includegraphics[width=4.5cm]{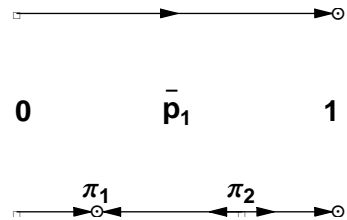}
\end{center}
\vspace{0.3cm}
\caption{ Schematic portrait of (\ref{21}) for $C=0$ [upper diagram] and for $C$
satisfying conditions (\ref{24}) [lower diagram]. Stable [unstable] rest points are denoted
by circles [squares]. Arrows indicate direction of flow in time.
}
\label{f_0}
\end{figure}

\section{Two morphs}
\label{two}

\subsection{Rest-points}

{

For $n=2$, Eq.~(\ref{z2}) simplifies to
a closed equation for the frequency $p_1$
\begin{gather}
\label{18}
\dot{p}_1(t)=p_1(1-p_1)[A(\tau)-B(\tau)p_1],\qquad \tau=\omega t,
\end{gather}
where we denoted for periodic functions of time [cf.~(\ref{corrida})]:
\begin{gather}
\label{simon1}
A(\tau)\equiv a_{12}(\tau)-a_{22}(\tau), \\
B(\tau)\equiv a_{21}(\tau)+a_{12}(\tau)-a_{11}(\tau)-a_{22}(\tau).
\label{simon2}
\end{gather}

\comment{
Before continuing, let us note that any non-trivial behaviour is
expected only when $A(t)$ change their sign in the course of time.
Otherwise, the sign of $\dot{p}_1$ will be fixed and we shall get the
same long-time behaviour as for constant $A$ and $B$. }

We proceed from (\ref{18}) along the lines of (\ref{z8}, \ref{9}). For the
frequency $p_1$ we get
\BEA
p_1=\bp_1+\frac{1}{\omega}\bp_1(1-\bp_1) \,[\,\hat{A}(\tau) - \hat{B}(\tau) \bp_1\,]+{\cal O}\left(\frac{1}{\omega^2}\right),
\EEA
where $\hat{A}(\tau)$ and $\hat{B}(\tau)$ are defined from $\widetilde{A}(\tau)$ and $\widetilde{B}(\tau)$
analogously to (\ref{6}, \ref{66}, \ref{6.1}). Now defining
\BEA
\label{u1}
C\equiv\frac{1}{\om}\, \overline{\hat{A}\,\widetilde{B}}
=\frac{1}{\om}\, \overline{  (\hat{a}_{12} - \hat{a}_{22} ) \,(\widetilde{a}_{21}-\widetilde{a}_{11}) }
\EEA
for the non-linear factor, we deduce from (\ref{z8}, \ref{9}, \ref{18}, \ref{darder})
\begin{gather}
\label{21}
\dot{\bp}_1(t)=\bp_1(1-\bp_1)[\bar{A}-\bar{B}\bp_1-C\bp_1(1-\bp_1)].
\end{gather}

The vertices $\bp_1=1$ and $\bp_1=0$ are always rest points of (\ref{21}), while
two interior rest points are
\BEA
\label{pipi}
\pi_{1,2}=\frac{1}{2C}[\bar{B}+C\mp \sqrt{(\bar{B}+C)^2-4\bar{A}C}\,\,],
\EEA
If $\pi_1$ and $\pi_2$ are in $(0,1)$, then
$\pi_1$ is stable, while $\pi_2$ is unstable. This follows from (\ref{pipi})
and also from the fact that in the vicinity of e.g. the rest-point $\pi_1$,
(\ref{21}) reads:
\begin{gather}
\label{210}
\frac{\d}{\d t}[{\bp}_1-\pi_1]=\pi_1(1-\pi_1)(\bp_1-\pi_1)(\bp_1-\pi_2)C.
\end{gather}
We also get $\pi_1<\pi_2$ for $C>0$ and $\pi_1>\pi_2$ for $C<0$.

The analysis of (\ref{21}) reduces to the following scenarios.

}

\subsection{Emergence of polymorphism}
\label{gorokh}

Let the time-averages hold
\BEA
\bar{A}>0 \quad {\rm and} \quad \bar{A}>\bar{B}.
\label{otch}
\EEA
Now the morph $1$ globally dominates for $C=0$, i.e., for all initial
conditions $\bp_1$ goes to $1$ for large times; see Fig.~\ref{f_0}.
($\bar{A}>0$ ensures that $p=0$ is unstable rest point, while
$\bar{A}>\bar{B}$ ensures that $p=1$ is a stable rest point.)
The global dominance does not change for $C<0$, because for a negative $C$
(and under conditions (\ref{otch})), both new stable point $\pi_1$ and
$\pi_2$ fall out of the interval $[0,1]$. One can call this dominating
morph generalist \cite{stein}, since it adapts to the time-averaged
environment.

\begin{figure}
\vspace{0.3cm}
\begin{center}
\includegraphics[width=7cm]{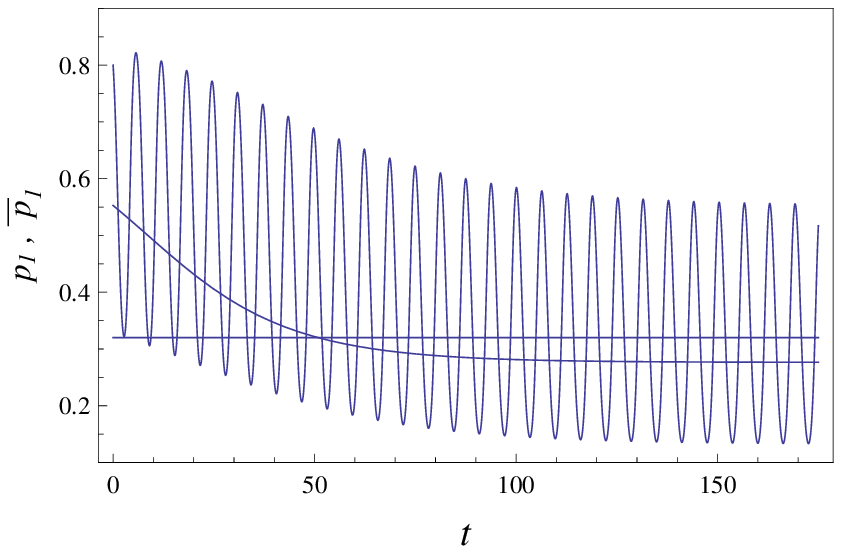}
\end{center}
\vspace{0.3cm}
\caption{\\ Oscillating curve: solution of (\ref{18}) with $A(\omega t)=0.1 - \sin (\omega t)$, $B(\omega t)=\cos (\omega t)$ with $\omega =1$ and the
initial condition $p_1(0)=0.8$. With these $A(\tau)$ and $B(\tau)$, the parameter $C$ in (\ref{u1}) is equal to $0.5$. \\
Smooth curve: solution of the effective equation (\ref{21}) with $\bar{A}=0.1$ and $\bar{B}=0$ and $C=0.5$, and initial condition
$\bp_1(0)=0.5528$. The difference between the initial conditions $\bp_1(0)$ and $p_1(0)$ is calculated according to (\ref{poncho}).
With time $\bp_1(t)$ converges to the rest point $\pi_1=0.276393$. For the unstable rest point we have $\pi_2=0.723607$.
\\
Straight line: the long-time average of $p_1(t)$ equal to $0.32$. The approximate equality between the long-time average and
$\pi_1$ improves upon increasing $C$ or decreasing $\bar{A}$ (and $|\bar{B}|$ if it is non-zero).
}
\label{ff_1}
\end{figure}

Provided that (\ref{otch}) holds, for
\BEA
\label{24}
C\geq 2\bar{A} - \bar{B}+2\sqrt{\bar{A}^{\, 2}-\bar{A}\bar{B}},
\EEA
i.e., when $C$ is large enough, both $\pi_1$ and $\pi_2$ fall into the
interval $[0,1]$, see Fig.~\ref{f_0}, while if condition (\ref{24}) does
not hold, both $\pi_1$ and $\pi_2$ are not in this interval.  When
condition (\ref{24}) saturates as equality, we get
\BEA
\pi_1=\pi_2=\frac{1}{\bar{B}}\, \left[\,\bar{A}-\sqrt{\bar{A}^{\,2}-\bar{A}\bar{B}}\,\right],
\nonumber
\EEA
which under conditions (\ref{otch}) is always in the interval $[0,1]$.
Thus, when $C$  increases from a smaller value and then starts to
satisfy (\ref{24}), $\pi_1$ and $\pi_2$ move from the complex plane into
the interval $[0,1]$, i.e., generically [from the viewpoint of
conditions (\ref{otch})] they do not appear in this interval via
crossing its boundaries. This is natural, because at the boundaries the
term $-C\bp_1(1-\bp_1)$ in (\ref{21}) nullifies.

Thus if condition (\ref{24}) holds [in addition to (\ref{otch})],
a stable rest point $\pi_1$ emerges, which attracts all the
trajectories that start from $\bp_1(0)<\pi_2$: the polymorphism is
created by the non-linear term $\propto C$ in (\ref{21}).

Initial condition larger than the unstable rest point $\pi_2$,
$\bp_1(0)>\pi_2$, still tend to $\bp_1=1$; see Fig.~\ref{f_0}. Both
stable rest points $\pi_1$ and $1$ are Evolutionary Stable States
(ESS), meaning that they cannot be invaded by a sufficiently small
mutant population \cite{hofbauer,zeeman}. The coexistence of two ESS one of
which is interior (i.e., polymorphic) is impossible for a two-player
replicator equation with constant pay-offs \cite{hofbauer,zeeman}, but it is
possible for multi-player replicator equation \cite{broom_vick}. We thus
saw above an example of this behavior induced by time-varying
environment.

\begin{figure}
\vspace{0.3cm}
\begin{center}
\includegraphics[width=7cm]{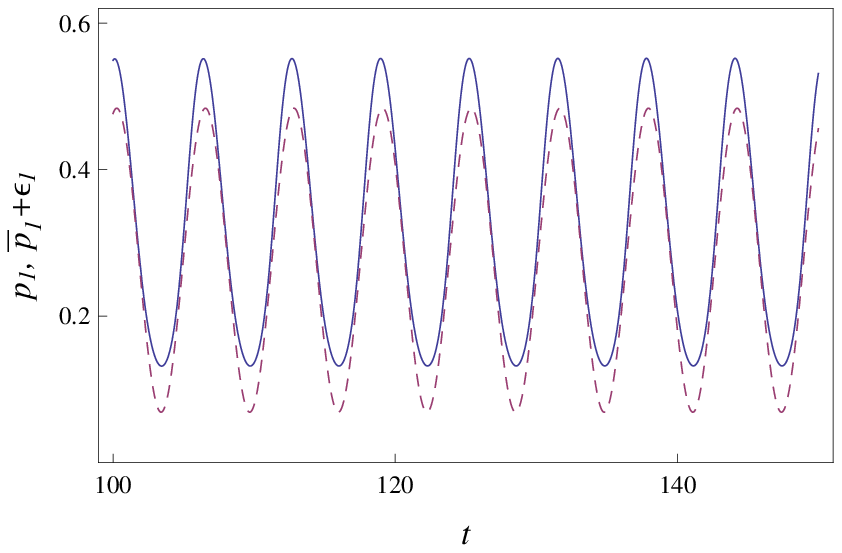}
\end{center}
\vspace{0.3cm}
\caption{Normal curve: $p_1(t)$ as obtained from solving (\ref{18}) 
with $A(\omega t)=0.1 - \sin (\omega t)$, 
$B(\omega t)=\cos (\omega t)$ with $\omega =1$,
and initial condition $p_1(0)=0.5$; cf.~Fig.~\ref{ff_1}. \\ Dashed
curve: $\bp_1(t)+\e_1(t)$, where $\bp_1(t)$ is obtained from solving the
effective equation (\ref{21}), and where $\e_1(t)=\e_1[\bp_1(t), \omega
t]$ is the corresponding oscillating factor found in (\ref{z8}). 
According to (\ref{21}), for considered times $\bp_1$ converged to the
stable fixed point $\pi_1$. For the present case $C=0.5$. 
The time-average of $p(t)$ does hold (\ref{21}). Comparing with 
Fig.~\ref{ff_23}, where $\bar{A}$ is larger, it is seen that
the approximate equality between $p_1(t)$ and $\bp_1(t)+\e_1(t)$ is better
for the smaller value of $\bar{A}$.
\label{ff_22}
}
\end{figure}

\begin{figure}
\vspace{0.3cm}
\begin{center}
\includegraphics[width=7cm]{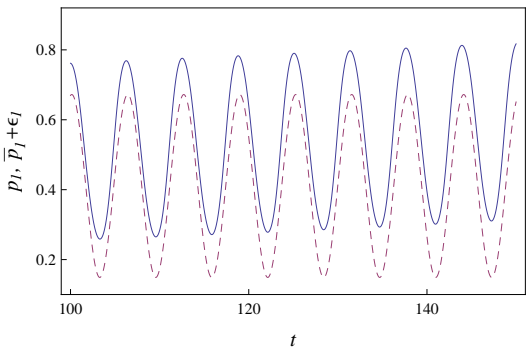}
\end{center}
\vspace{0.3cm}
\caption{ The same as in Fig.~\ref{ff_22}, but with $A(\omega t)=0.121 - \sin (\omega
t)$, i.e. with a larger value of $\bar{A}$. The figure represents a
metastable polymorphism, since $p(t)$ tends to increase slowly, i.e. it
does not hold (\ref{21}) for long times, and coverges to $p_1=1$ for times $t>230$.
\label{ff_23}
}
\end{figure}

\subsection{Modification of the existing polymorphism }
\label{strabon}

For $C=0$ and
\BEA
\label{schiller}
\bar{B}>\bar{A}>0 
\EEA
there is a stable polymorphism at the rest point
$\bp_1={\bar{A}}/{\bar{B}}$.  The presence of $C\not =0$ in (\ref{21})
does not change this polymorphism qualitatively; only the value of the
rest point shifts to $\pi_1$, which is now always in the interval
$[0,1]$.  The shift of the fixed point can be sizable.

\begin{figure}
\vspace{0.3cm}
\begin{center}
\includegraphics[width=7cm]{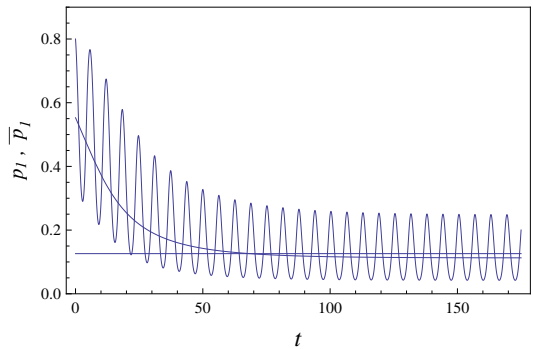}
\end{center}
\vspace{0.3cm}
\caption{ As compared to Fig.~(\ref{ff_1}) the value of $\bar{A}$ is decreased and
is now equal to $0.05$. This changes the values of the rest points:
$\pi_1=0.112702$ and $\pi_2=0.887298$. $p_1(t)$ and $\bp_1(t)$ are still
given by, respectively, oscillating curve and smooth curve.  The value
of the long-time average of $p_1(t)$ is now $0.1261$ (straight line),
which is rather close to $\pi_1$ (the long-time limit of the smooth
curve). }
\label{ff_3}
\end{figure}

Likewise, for
\BEA
\label{prudon3}
\bar{B}<\bar{A}<0,
\EEA
and $C=0$ there is an unstable polymorphism: all
the initial conditions with $\bp_1(0)<{\bar{A}}/{\bar{B}}$ end up at
$p_1=0$ (morph 2 dominates), while those with
$p_1(0)>{\bar{A}}/{\bar{B}}$ finish at $p_1=1$ (morph 1 dominates).  Now
$C\not =0$ in (\ref{21}) shifts the unstable rest point to $\pi_2$.
Again, the shift can be significant in some situations.

Conditions (\ref{otch}, \ref{24}, \ref{schiller}, \ref{prudon3}) gave all possible
non-trivial scenarios for two morphs, because other relations between
$\bar{A}$ and $\bar{B}$ amount to interchanging the morphs.

\subsection{Validity of the theory beyond the large-$\om$ assumption}

We now proceed to numerical comparison between the exact equation
(\ref{18}) and the effective (time-averaged) equation (\ref{21}).  Our
aim is to show that the approximate validity of (\ref{21}) extends beyond
the large-$\omega$ limit. 

Figs.~\ref{ff_1}, \ref{ff_22} and \ref{ff_23} compare the solution $p_1(t)$ of
(\ref{18}) with the corresponding solution $\bp_1(t)$ of (\ref{21}).
Here we took $\omega=1$ to make clear that the frequency is not very
large, and hence we are looking beyond the applicability of the above 
theory.  The initial conditions $p_1(0)$ and $\bp_1(0)$ were adjusted
acording to (\ref{poncho}). It is seen that though the qualitative
agreement is good|in particular, $p_1(t)$ does converge with time to a
polymorphic state, where its time-average is constant| there is a
visible quantitative difference between the long-time averages
calculated from the solution (\ref{18}) and the rest point $\pi_1$
predicted from (\ref{21}).  The quantitative agreement is improved
either by increasing $\omega$ (i.e., making the environmental
oscillations faster) or by decreasing $\bar{A}$ and $\bar{B}$ (i.e., by
slowing down the change of $p_1(t)$). This is demonstrated in
Fig.~\ref{ff_3}. 

\begin{figure}
\vspace{0.3cm}
\begin{center}
\includegraphics[width=7cm]{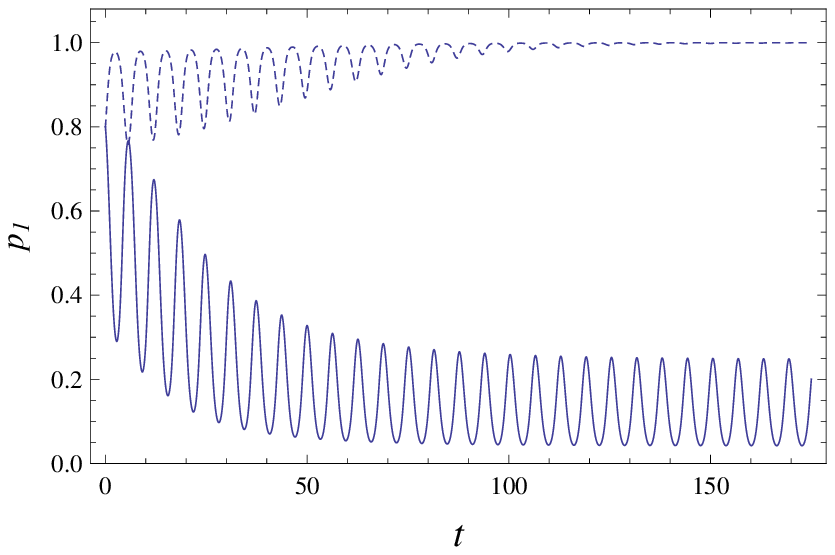}
\end{center}
\vspace{0.3cm}
\caption{ This figure illustrates how the basins of attraction for two stable rest points $\pi_1$ and $1$ change depending
on the phase of oscillating functions $A(\tau)$ [$\tau=\omega t$] and $B(\tau)$ at the initial time $t=0$. \\
Normal curve: solution of (\ref{18}) with $A(\tau)=0.1 - \sin (\omega t)$, $B(\tau)=\cos (\omega t)$ with $\omega =1$ and the
initial condition $p_1(0)=0.8$. \\
Dashed curve: solution of (\ref{18}) with $A(\tau)=0.1 + \sin (\omega t)$, $B(\tau)=-\cos (\omega t)$ with $\omega =1$ and the same
initial condition $p_1(0)=0.8$. \\
It is seen that for one phase $p_1(t)$ converges in averages to $\pi_1$, while for another phase it goes to $1$.
}
\label{ff_4}
\end{figure}

As discussed in section \ref{inici}, the same initial conditions
$p_1(0)$ can lead to a different attractor depending on the initial
phase of the oscillating functions $A(\tau)$ and $B(\tau)$. This phenomenon is
clearly seen on Fig.~\ref{ff_4}. We stress that this dependence is
not a special feature of the considered stable polymorphic state. It
will appear whenever there is a fast environment and the space of $p_1$
posseses several attractors, e.g., for the unstable polymorphic case
studied in section \ref{strabon}; see (\ref{prudon3}).

\begin{figure}
\vspace{0.3cm}
\begin{center}
\includegraphics[width=7cm]{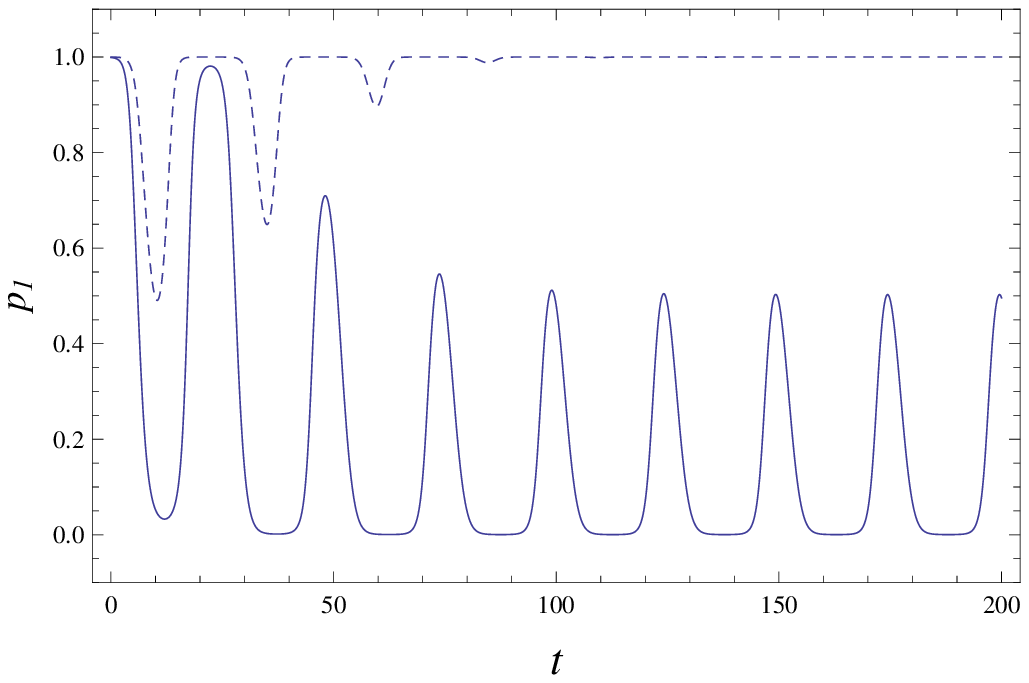}
\end{center}
\vspace{0.3cm}
\caption{
The purpose of this figure is to illustrate that the qualitative picture of polymorphism survives for smaller values of $\omega$,
i.e., it survives for slower environment.\\
The same parameters as in Fig.~\ref{ff_1}, besides $\omega=0.25$.\\
Normal curve: solution of (\ref{18}) with
initial condition $p_1(0)=0.999$.\\
Dashed curve: solution of (\ref{18}) with
initial condition $p_1(0)=0.9999$.\\
It is seen that the stability domain of the rest point $p_1=1$ is heavily reduced.
}
\label{ff_5}
\end{figure}

Fig.~\ref{ff_5} shows that polymorphism survivies for a smaller
frequency $\omega$. 

Fig.~\ref{ff_23} demonstrates an interesting phenomenon
of metastable polymorphism: the correspondence between (\ref{18}) and
(\ref{21})| i.e. between $p_1(t)$ and $\bp_1+\epsilon_1$| is reasonable
for times up $t\simeq 200$ (this characteristic time depends on initial 
conditions, the estimate seen in Fig.~\ref{ff_23} was got with initial 
conditions $p_1(0)=0.5$). For larger times, the real solution
converges to $p_1=1$. Thus $\bp_1+\epsilon_1$ is capable of describing 
the metastable polymorphism within its life-time that is much longer than
the characteristic time $1/\omega$ of environmental oscillations.

\comment{
We recognize that for such slow environments the
frequency $p_1$ approaches too closely to $1$ and/or to $0$, which means
that various mechanisms not accounted for by (\ref{18})|e.g., mutation,
stochastic effects leading to extinction, effects of finite
populations|will be important. This is why we do not dwell in greater
detail on the consideration of slow environments. Still it is
reassuring to see that qualitative predictions gained from 
(\ref{21}) survive also for slow environments. }

{

\subsection{Two morphs within mathematical genetics}

\subsubsection{Interpretations in terms of genetics}

Eq.~(\ref{18}) under additional condition
\BEA
a_{12}(\tau)=a_{21}(\tau).
\label{fisher_holden}
\EEA
is a well-known Fisher
equation from population genetics that describes the selection for a
two-alelle gene (with frequencies $p_1$ and $p_2$, respectively) in one
locus \cite{svi,nagylaki_book}. Then $a_{kl}$ are the Malthusian
fitnesses for the zygote formed by the alleles $k$ and $l$. In
particular, $a_{11}$ and $a_{22}$ refer to two homozygotes, while
$a_{12}=a_{21}$ refers to the heterozygote. (A short and accessible
review on genetic notions is given in Ref.~\cite{svi}.)

Let us now briefly comment on some of above results in the light of
mathematical genetics, i.e. assuming (\ref{fisher_holden}). {\it (1)}
Conditions (\ref{schiller}) for a stable polymorphism translate to
$\bar{a}_{12}>\bar{a}_{11},\, \bar{a}_{22}$, and refer to the
heterozygote advantage \cite{svi}. 

{\it (2)} Note that $C$ in (\ref{u1}, \ref{21}) vanishes whenever the
homozygotes are symmetric for all times ${a}_{11}(\tau)={a}_{22}(\tau)$,
or whenever one allele (say allele 2) is recessive for all times
${a}_{11}(\tau)={a}_{12}(\tau)$. Both cases are easily solvable from
(\ref{18}) showing that the long-time behavior of $p_1$ is indeed
governed by $\bar{A}$; so there is no room for the non-linear terms. 

{\it (3)} The simplest case for non-zero $C$ is perhaps when the
heterozygote fitness $a_{12}$ is constant in time, but the homozygote
fitnesses $a_{11}(t)$ and $a_{22}(t)$ oscillate in time with different
phases, e.g., when $a_{11}(t)$ is maximal $a_{22}(t)$ is minimal and
{\it vice versa}. 

{\it (4)} In section \ref{strabon}, after (\ref{schiller}) 
we stressed that the environmental influence, i.e. $C\not=0$, 
can sizably shift the stable rest-point thereby facilitating polymorphism. 
An important example of this type is given by the following
mechanism of the recessive allele survival. For
\BEA
\label{ss}
\bar{B}=\bar{A}>0 \quad {\rm or}\quad \bar{a}_{12}=\bar{a}_{11}>\bar{a}_{22},
\EEA
the time-averaged fitnesses predict that the allele $2$ is recessive,
i.e., its presence does not influence on the fitness of the
heterozygote, while the fitness of the corresponding homozygote $22$ is
smaller.

For $C<\bar{A}$ (in particular for $C=0$) the only stable rest point is $p_1=1$, which means that the allele is absent from the
population. However, for $C>\bar{A}$ the poymorphism is recovered,
since now [see (\ref{pipi})]
\BEA
\pi_1=\bar{A}/C.
\EEA
Thus the rapidly-changing environment can lead to a significant expression
of the allele which is recessive in average.

\subsubsection{Previous literature}

Periodic time-dependencies in $A(t)$ and $B(t)$ [cf.~(\ref{18})] were
studied by Kimura \cite{kimura}, Nagylaki \cite{nagylaki}, and Li
\cite{li} in modelling environmental influences on genetic selection.
These authors concentrate on cases (e.g. $A(t)=0$ or $B(t)=0$), where
non-trivial mechanisms of polymorphism are absent. Below we shall focus
on those situations, where Eq.~(\ref{18}) with time-independent $A$ and
$B$ is not solvable exactly, but instead we obtain a non-trivial
scenario of polymorphism. 

Within the continuous-time consideration,
Nagylaki \cite{nagylaki} focussed on the case 
[cf.~(\ref{mashad})]
\BEA
\label{retro}
A(\tau)=a\,h(\tau), \qquad B(\tau)=b\, h(\tau), \qquad \tau=\omega t
\EEA
where $h(\tau)$ is a periodic function with $\overline{\,h(t)\,}=0$, and
where $a$ and $b$ are constants. Note that $\bar{A}=\bar{B}=0$ according
to (\ref{retro}). Now (\ref{18}) can be solved exactly (i.e. independently 
from the magnitude of $\omega$), leading one to
conclude that (\ref{18}) cannot have stable rest-points, i.e. the
convergence in time $p_1(t)\to p_{\rm rp}$ is excluded for $p_{\rm rp}\in
[0,1]$. We are however interested in stable rest-points for the
time-average $\bp$, hence we turn to analyzing (\ref{21}) under (\ref{retro})
and a sufficiently large $\omega$. 

For this case we obtain from (\ref{21}): $\dot{\bp}_1=0$, a conclusion
that is very well confirmed numerically and extends to higher orders in
the $\frac{1}{\om}$ expansion; see Appendix \ref{mansur} and (\ref{ushi}) there.  Hence
Nagylaki did not find polymorphism scenarios under conditions
(\ref{retro}). 

}

\subsection{Average fitness and Lyapunov function}

Eqs.~(\ref{kondor}) for the mean, time-averaged 
fitness $\Phi$ read after using (\ref{simon1}, \ref{simon2})
\BEA
\label{golan}
\Phi=2\bar{A}\bp_1-\bar{B}\bp_1^2.
\EEA

For $\bar{A}>0$ and $\bar{A}>\bar{B}$ (where the first morph dominates
according to the average fitness), we get $\partial\Phi/\partial \bp_1>0$. Hence
$\Phi$ maximizes at $\bp_1=1$, and
this maximum is the only stable rest point of the replicator dynamics
(\ref{21}) with $C=0$. If however $C$ satisfies conditions (\ref{24}),
in the stable rest point $\bp_1=\pi_1$ the mean fitness $\Phi$ is
smaller than at the stable point $\bp_1=1$. Moreover, for the initial
conditions $\pi_1<\bp_1(0)<\pi_2$, the mean fitness $\Phi$ decreases in
the course of the relaxation to $\pi_1$. We confirmed this 
statement directly from equations of motion; see Table~I.

Hence, the existence of a polymorphism does not have an adaptive value
for the overall population, because the overall fitness decreases. We
saw in section \ref{strabon} that the polymorphism which already exist
on the level of the average environment appears to be robust with
respect to environmental variations. So the value of polymorphism might
be connected with this robustness. 

Eq.~(\ref{21}) can be written as
\BEA
\dot{\bp}_1&=&\bp(1-\bp)\,\frac{\partial \Psi}{\partial \bp},\\
\Psi&=&2\bar{A}\bp_1-\bar{B}\bp_1^2
-C\bp_1^2(1-\frac{2\bp_1}{3})\nonumber \\
&=&\Phi-C\bp_1^2(1-\frac{2\bp_1}{3}).
\EEA
Hence $\Psi$ is increased by dynamics (\ref{21}): $\dot{\Psi}\geq 0$.
The difference between the mean fitness $\Phi$ and the Lyapunov function $\Psi$
is yet another indication that $\Phi$ does not need to increase in the slow time. 
Let us emphasize that there is no relation between the sign of $\Psi-\Phi$
and the existence of the environment-induced polymorphism.

\begin{table} \caption{The solution of (\ref{18}) was obtained under the
same parameters as for Fig.~\ref{ff_1}. Then the time-dependence overall
fitness $\Phi(t)$ was calculated accroding to
$\Phi(t)=2A(\tau)p(t)-B(\tau)p^2(t)$; see (\ref{kondor}, \ref{simon1},
\ref{simon2}) and note an irrelevant factor $a_{22}(t)$ in $\Phi(t)$ was
neglected. After that we time-averaged this fitness around the points
$t_0$ with the same time-window $t_w=25$: $\bar{\Phi}(t_0)=
\frac{1}{2t_w}\int_{t_0-t_w}^{t_0+t_w} \d s\, \Phi(s) $. The results for
$\bar{\Phi}(t_0)$ are displayed below and show that the time-averaged
fitness decreases with increasing $t_0$.  } \begin{tabular}{|c||c|}
\hline\hline
     ~$t_0 = 25 $~ & ~$0.09627$~ \\
      \hline
     ~$t_0 = 75 $~ & ~$0.07192$~  \\
      \hline
     ~$t_0 = 125$~ & ~$0.06467$~  \\
      \hline
     ~$t_0 = 175$~ & ~$0.06244$~  \\
      \hline
     ~$t_0 = 225$~ & ~$0.06167$~  \\
\hline\hline
\end{tabular}
\label{tab_1}
\end{table}

\section{The prisoner's dilemma}
\label{remind}

As another pertinent illustration we consider 
the prisoner's game \cite{shubik}.
There are two players $I$ and $II$. Each one has two strategies:
$d$ (defect) and $c$ (cooperate). Pay-offs are determined by the following matrix
\begin{eqnarray}
\label{bad}
\begin{tabular}{|c|c|c|}
  \hline
~$ {I}/{II}$~  & $d$ & $c$ \\
  \hline
  ~$d$~ & ~$P, \, P$~ & ~$T, \, S$~ \\
  \hline\hline
  ~$c$~ & ~$S, \, T$~ & ~$R, \, R$~ \\
  \hline
\end{tabular}
=\begin{tabular}{|c|c|c|}
  \hline
~$ {I}/{II}$  & $1$ & $2$ \\
  \hline
  $1$ & $a_{11}, a_{11}$ & $a_{12}, a_{21}$ \\
  \hline\hline
  $2$ & $a_{21}, a_{12}$ & $a_{22}, a_{22}$ \\
  \hline
\end{tabular}\, ,
\end{eqnarray}
where e.g. the actions $d$ and $c$ by respectively $I$ and $II$ result
to pay-offs $T$ and $S$, and where the second matrix in (\ref{bad}) 
relates to (\ref{z1}, \ref{z2}). Eq.~(\ref{bad}) becomes
a dilemma after imposing:
\begin{eqnarray}
\label{pri}
T>R>P>S.
\end{eqnarray}
For both players defecting ($d$) is a dominant strategy, i.e.  for both
$I$ and $II$ acting $d$ yields a higher payoff than $c$ (cooperating),
no matter what the opponent does; see (\ref{bad}, \ref{pri}) and note
that $(d,d)$ is the only Nash equilibrium of game (\ref{bad}).  
Both players can get $R>P$, if they both act $c$. But
acting $c$ is vulnerable, since the opponent can change to $d$,
gain out of this, and leave the cooperator with the minimal pay-off $S$.
This makes the dilemma, which raised deep questions about rationality
and cooperation \cite{shubik,hofbauer,peterson} and produced a
vast literature 
\cite{sandler,dyson,15saakian,weitz,benica}. 

We focus on the case when pay-offs in (\ref{bad}) are time-periodic
functions. Starting from (\ref{z1}, \ref{z2}, \ref{corrida}),
the time-dependent replicator equation reduces to (\ref{18}),
where 
\begin{align}
\label{pr1}
&A(\tau)=T(\tau)-R(\tau),\qquad \tau=\omega t,\\
&B(t)=T(\tau)-R(\tau)+S(\tau)-P(\tau).
\label{pr11}
\end{align}
If (\ref{pri}) is imposed on pay-offs at all times, then $A(\tau)>0$ and
$A(\tau)>B(\tau)$. Then the always defection strategy $p_1=1$ is the only
rest-point of the time-dependent (\ref{18}), i.e. we are back to the
prisoner's dilemma. 

{

A reasonable way to study the prisoner's game for a rapidly changing environment
is to impose condition (\ref{pri}) only on average
\BEA
\label{priav}
\bar T>\bar R>\bar P>\bar S.
\EEA
Now (\ref{pr1}--\ref{priav}) show that we are in the situation discussed
in section \ref{gorokh}: there is no polymorphism on the level of the
average fitness. But if condition (\ref{24}) holds, the polymorphism 
does emerge. The positivity of $C=\frac{1}{\om}\overline{\, 
(\hat T-\hat R)(\widetilde S-\widetilde P) \,}$ means here that the 
changes of $\widetilde T(\tau)-\widetilde R(\tau)$ 
and of $\widetilde S(\tau)- \widetilde P(\tau)$ are correlated. 

Hence a time-dependent environment can resolve the prisoner's dilemma
that exists on the level of the average fitness. Other scenarios of
resolving the dilemma are reviewed in \cite{tanimoto,helbing,devlin}. 

}

\section{Three morphs}
\label{three}

\subsection{Rock-scissors-paper game}

Besides mathematical genetics \cite{nagylaki_book,svi}, the replicator
dynamics is also applied for modeling biodiversity
\cite{hofbauer,zeeman}. For concretness we shall focus on the situation
with cyclic dominance, which exists for at least three morphs and
requires asymmetric pay-offs. Cyclic dominance means that morphs $1$,
$2$ and $3$ win over each other; e.g. $1$ beats $2$, $2$ beats $3$, and
$3$ beats $1$ (rock-scissors-paper game) \cite{abraham}. The simplest
and perhaps most popular example of cyclic dominance is realized under
the zero-sum condition in (\ref{z2}) [see (\ref{51}) for an example]:
\BEA
\label{250}
a_{kl}(\tau)=-a_{lk}(\tau), \qquad \tau=\om t.
\EEA
Eq.~(\ref{250}) means that the loss of the strategy $l$ is equal to the
gain of $k$.  Here is an incomplete list of realistic examples that
contain cyclic dominance: {\it i)} mating strategies of side-blotched
lizards \cite{sinervo}; {\it ii)} overgrowths by marine sessile
organisms \cite{buss}; {\it iii)} competition between bacterial
populations \cite{kerr}.

\comment{(ready part)
, where there are three different strains
(morphs) : toxin producer (1), toxin sensitive (2) and toxin resistant
(3).  Now 1 beats (poisons) 2, 3 beats 1 because it is not sensitive to
the poison, and does not pay metabolic costs for toxin production, and
finally 2 beats 3, because 2 does not pay immunization costs.
Note that the 1-3 and 2-3 couplings are due to resource competition.}

\begin{figure}
\vspace{0.3cm}
\begin{center}
\includegraphics[width=7cm]{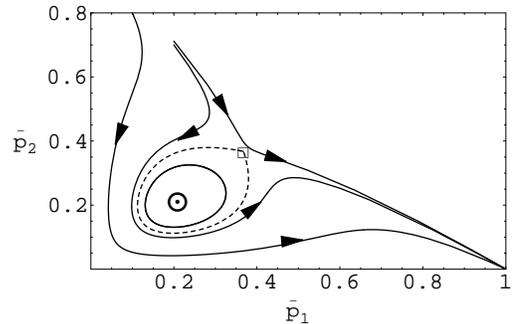}
\end{center}
\vspace{0.3cm}
\caption{ The portrait of (\ref{31}) for $\bar{a}_{12} =0.1$, $\bar{a}_{13} = 0.1$, $\bar{a}_{23} = -0.1$, $\kappa_1 = -0.65$ and
$\kappa_2 = 0.65$; $\bar{p}_1$ and $\bar{p}_2$ are restricted by $0\leq
\bar{p}_1\leq 1$, $0\leq \bar{p}_2\leq 1$, $\bar{p}_1+\bar{p}_2\leq 1$. Two rest points
are denoted by square (saddle) and cycle (center).
The closed orbits contain the center in
their interior; orbits from the second class converge to $\bar{p}_1=1$. These
two classes of orbits are separated by a dashed curve (separatrix), which is made by joining
together two directions (stable and unstable) of the saddle. Arrows indicate direction of flow.
}
\label{f_1}
\end{figure}

Though the zero-sum condition frequently does not produce structurally stable results
\footnote{A structurally stable model for the oscillating regime of the
rock-paper-scissors game has been recently proposed in \cite{schuster}.}, it is still very
useful, since one can give a general description of the constant-payoff,
$a_{kl}(\tau)=\bar{a}_{kl}$, zero-sum situation \cite{akin}: starting
from the interior of the simplex|that is starting from a point, where
all the fractions are strictly positive: $p_k>0$|any trajectory either
remains in the interior performing there a Hamiltonian motion
\footnote{A Hamiltonian motion means that there is one globally
conserved function of the fractions $p_k(t)$, and that the suitably
defined phase-space volume is conserved; see \cite{akin} for details. },
or it converges to the boundary, where some fractions $p_k$ are equal to
zero. Once restricted to the boundary, the situation reduces to a
zero-sum game with a smaller number of variables and the above reasoning
can be applied again \cite{akin}.  Which scenario is realized depends on
the concrete form of $a_{kl}(\tau)=\bar{a}_{kl}$; see below.

We stress that we also studied the three-morph case for the symmetric
pay-off situation $a_{kl}(\tau)=a_{lk}(\tau)$, but the qualitative
results on the environment-induced polymorphism were similar to the
zero-sum case [see in this context section \ref{Poincare} below], so we
choose to concentrate on the latter, because its presentation is
simpler.

Under condition (\ref{250}) the effective replicator
equation (\ref{11}, \ref{12}) reduces to
\BEA
\label{31}
\dot{\bp}_i={\sum}_{\alpha=1}^3 \bp_i \bar{a}_{i\alpha}\bp_\alpha +\kappa_i
\bp_1\bp_2\bp_3,\qquad i=1,2,3,
\EEA
where
\BEA
\label{34}
\kappa_1=b_{123}+b_{132}=\frac{1}{\om}(-\overline{\widetilde{a}_{13}\,\hat{a}_{23}}+
\overline{\widetilde{a}_{12}\,\hat{a}_{23}}),\\
\label{35}
\kappa_2=b_{213}+b_{231}=-\frac{1}{\om}(\overline{\widetilde{a}_{23}\,\hat{a}_{13}}+
\overline{\widetilde{a}_{12}\,\hat{a}_{13}}),
\EEA
with
\BEA
\label{36}
\kappa_1+\kappa_2+\kappa_3=0,
\EEA
as needed for the conservation of normalization.  It is seen that for
the zero-sum situation the four-party terms $c_{klmn}$ disappear, since
the payoffs are anti-symmetric.

Recall the interpretation of (\ref{31}) with $i=1$: besides the payoffs $a_{12}$
and $a_{13}$ got by the strategy 1 when confronting with 2 and 3,
respectively, there is a payment $\kappa_1$ which is received by the
strategy 1 when it is confronted with the strategies $2$ and $3$
together. Eqs.~(\ref{31}) $i=2,3$ are interpreted in a similar way.

\begin{figure}
\vspace{0.3cm}
\begin{center}
\includegraphics[width=7cm]{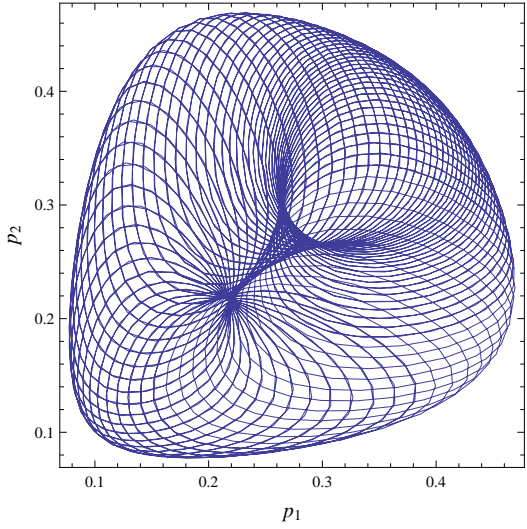}
\end{center}
\vspace{0.3cm}
\caption{
Parametric plot of solutions $p_1$ and $p_2$ of the replicator 
equation (\ref{z1}, \ref{z2}) for $n=3$ (three morphs),
$t\in [0, 600]$ and $a_{12}(\om t)=-a_{21}(\om t)=0.1$, $a_{13}(\om t)
=-a_{31}(\om t)=0.1+\sqrt{2\om}\,a \sin(\om t)$,
$a_{23}(\om t)=-a_{32}(\om t)=0.1+\sqrt{2\om}\,a \cos(\om t)$, $a=\sqrt{0.65}$,
$\om=2$. The initial conditions are $p_1(0)=p_2(0)=0.25$.\\
Then the effective motion for $\bp_1$ and $\bp_2$ holds
parameters shown on Fig.~\ref{f_1} independently on the value of $\om$. 
}
\label{f_8}
\end{figure}

\begin{figure}
\vspace{0.3cm}
\begin{center}
\includegraphics[width=7cm]{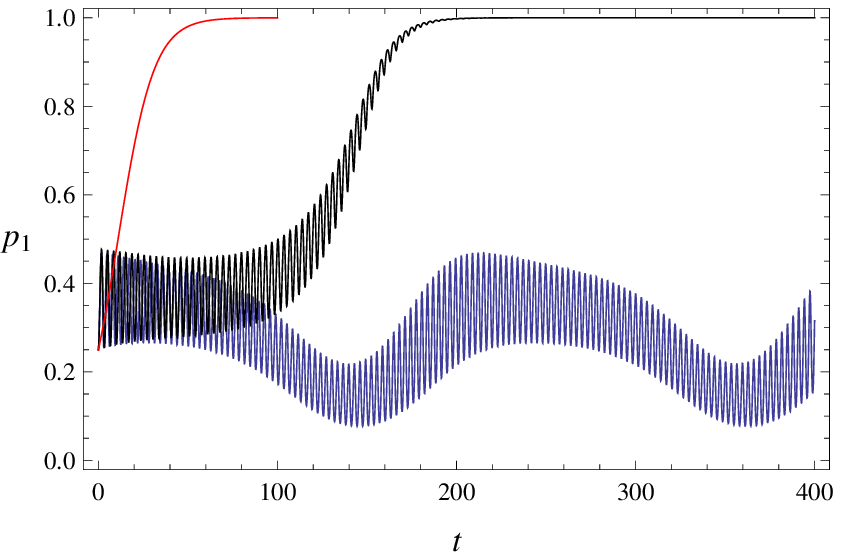}
\end{center}
\vspace{0.3cm}
\caption{The first morph population $p_1$ versus time for the same parameters as in Fig.~\ref{f_8}. \\
Black curve: $\om=1.85$. Blue curve: $\om=2$. Red curve: no environmental changes: $\om=0$. \\
It is seen that $p_1(t)$ performs fast oscillations, and then slower motion that is oscillatory 
for the blue curve and increasing for the black curve.
}
\label{f_9}
\end{figure}

The phase structure of (\ref{31}) will be constructed in
terms of independent variables $\bp_1$ and $\bp_2$ with $0\leq \bp_1+\bp_2\leq
1$. For the internal rest points we have:
\begin{align}
&\ba_{12} \bp_2+\ba_{13}(1-\bp_1- \bp_2)+\kappa_1 \bp_2(1-\bp_1- \bp_2)=0,\nonumber\\
-&\ba_{12} \bp_1+\ba_{23}(1-\bp_1- \bp_2)+\kappa_2 \bp_1(1-\bp_1- \bp_2)=0.
\end{align}
The Jacobian at these rest points is
\BEA
-\left(\begin{array}{rr}
\bp_1(\ba_{13}+\kappa_1 \bp_2) ~~~& \frac{\bp_1}{\bp_2}(\ba_{13}(1-p_1)+\kappa_1\bp_2^2) \\
\\
\frac{\bp_2}{\bp_1}(\ba_{23}(1-\bp_2)+\kappa_2\bp_1^2) & -\bp_1(\ba_{13}+\kappa_1 \bp_2)~~~ \\
\end{array}\right),\nonumber
\EEA
while the eigenvalues of the Jacobian reads:
\begin{align}
&\pm\left[ \,\bp_1^2(\ba_{13}+\kappa_1 \bp_2)^2  \right.\nonumber\\
&\left. +(\ba_{13}(1-\bp_1)+\kappa_1\bp_2^2)(\ba_{23}(1-\bp_2)+\kappa_2\bp_1^2)   \right]^{1/2}.
\end{align}
Thus any interior rest point can be either saddle (two real eigenvalues
of different sign) or center (two imaginary, complex conjugate
eigenvalues).

We study in separate two different cases.

{

\subsection{Emergent polymorphism}

One morph (say 1) dominates at the level of average pay-offs:
\BEA
\label{50}
\ba_{12}>0, \quad \ba_{13}>0.
\EEA
Eq.~(\ref{31}) with $i=1$ shows that the dominance is kept for
$\kappa_1>0$.  Hence for the existence of a polymorphism (due to
$\kappa_i$ terms) it is necessary that 2 and 3 {\it together} win over 1
($\kappa_1<0$), although {\it in separate} they loose to 1 according to
(\ref{50}). 

There are two further necessary conditions for polymorphism that are
deduced from the requirement that terms $\kappa_i\bp_1\bp_2\bp_3$ do not
turn to zero due to $p_2\to 0$ and/or $p_3\to 0$. Hence, as (\ref{31})
shows, we should exclude the following two conditions:
\BEA
&&\ba_{23}<0 \quad {\rm and} \quad \kappa_2<0,\\
&&\ba_{23}>0 \quad {\rm and} \quad \kappa_3=-\kappa_1-\kappa_2<0.
\EEA
Then it is possible that the (non-linear) terms $\propto
\kappa_1$, $\propto \kappa_2$ and $\propto \kappa_3$ in
(\ref{31}) change the phase-space portrait. Due to
the non-linear terms two interior rest points appear in the triangle
$0\leq \bp_1\leq 1$, $0\leq \bp_2\leq 1$, $\bp_1+\bp_2\leq 1$; see
Fig.~\ref{f_1}. The rest point with a smaller $\bp_1$ is center, while
the rest point with a larger $\bp_1$ is a saddle.

The phase-space is thus divided into two domains: the first domains
contains closed orbits around the center; see Fig.~\ref{f_1}.  These
orbits correspond to the polymorphism: all the fractions $\bp_1$, $\bp_2$
and $\bp_3$ stay non-zero for all times.

The second domain contains trajectories that converge to $\bp_1=1$ for
large times; see Fig.~\ref{f_1}. The two domains are divided by the
separatrix, a closed orbit that passes through the saddle; see
Fig.~\ref{f_1}. This scenario of polymorphism is similar to the one
for the two-morph situation; see around (\ref{24}).

Figs.~\ref{f_8} and \ref{f_9} show the behaviour of the original
replicator equation (\ref{z1}, \ref{z2}) with asymmetric pay-offs
[cf.~(\ref{250})], whose time-averages relate to (effective) parameters of
Fig.~\ref{f_1}. Now Fig.~\ref{f_8} shows that a single trajectory fills
a sizable part of the phase-space available for the polymorphism.
Fig.~\ref{f_9} contrasts the behavior of $p_1(t)$ in the polymorphic
regime|which consists of fast and slow oscillations|with what we called
a metastable polymorphism in Fig.~\ref{ff_23}; see the black
curve in Fig.~\ref{f_9}. Now the metastable polymorphism is realized
when the frequency of environmental oscillations is not sufficiently
large, i.e. $\om=1.85$ versus $\om=2$ for parameters of Figs.~\ref{f_8}
and \ref{f_9}. It amounts to fast oscillations of $p_1(t)$ around a
value sizably smaller than $1$, which then squeeze (for $t>100$), and
then $p_1(t)$ goes to $1$, i.e. the polymorphism is destroyed, since the
morph 1 eventually wins over $2$ and $3$; see the black curve on
Fig.~\ref{f_9}. Note that the life-time, $t\simeq 100$, of this
metastable polymorphism is much larger than the period of environmental
oscillations, as well as the relaxation time (to $p_1=1$) for the
time-independent environment; see Fig.~\ref{f_9} and cf.~Fig.~\ref{ff_23}.

\subsection{Existing polymorphism}

Now let us assume that there is polymorphism already at the average level:
\BEA
\label{51}
\ba_{12}>0, \quad \ba_{13}<0, \quad \ba_{23}>0,
\EEA
which means a cyclic competition: the strategy 1 wins over 2, but looses
to 3, while 2 wins over $3$.  This is the zero-sum version of the
rock-scissors-paper game \cite{hofbauer,zeeman}.

Thus for $\kappa_1=\kappa_2=\kappa_3=0$ in (\ref{31}) there is
already one rest interior point, and the trajectories of the system are
closed orbits around this rest point. It appears that after including
the three-party ($\propto \kappa_i$) terms in (\ref{31}) this
rest point is simply shifted, and no new rest points appear for any size
or magnitude of $\kappa_i$. We get the same conclusion as for the $n=2$
case: the non-linear (multi-party) terms do not spoil the polymorphism
that already exists without them; see section \ref{strabon}.

}

\section{Features of polymorphism related to Poincar\'e indices}
\label{Poincare}

In sections \ref{two} and \ref{three} we noted that the non-linear terms
(\ref{14}) in the effective replicator equation produce either two new
rest points (one stable and another unstable), or do not produce new
rest points at all, although they can sizably shift the existing
interior rest points (interior means that all components of the rest
point are positive). Moreover, generically the new fixed points are
produced directly in the interior, i.e., they do not have to appear via
crossing of the boundaries for the simplex region $\{0\leq \bp_k\leq 1
\}_{k=1}^n$, $\sum_{k=1}^n \bp_k=1$.

These effects suggest a common mechanism, which will be discussed below
in terms of Poincar\'e indices. Recall that rest-points of the
(non-linear) replicator dynamics (\ref{11}, \ref{12}) are defined via
$\Gamma_k[\pi]=0$.  At regular rest points the Jacobian has non-zero
determinant. We do not consider non-regular rest points, because they
turn to be regular under small perturbation.  Each regular rest point
$\pi$ can be associated with its Poincar\'e index:
\BEA
\label{index}
{\rm Ind}[\pi]={\rm sign}\,\,{\rm det}\left.\left( \frac{\partial \Gamma_i[\bp]}{\partial \bp_k}  \right)\right|_{\bp=\pi},~~
i,k=1,\ldots,n-1,\nonumber
\EEA
where $\frac{\partial \Gamma_i[\bp]}{\partial \bp_k}$ is the Jacobian
matrix, and where we assume that in defining the Jacobian an independent
set of coordinates was selected (since $\sum_{k=1}^n\bp_k=1$, only $n-1$
probabilities $\bp_k$ are independent). For instance in the
two-dimensional situation (i.e., for two independent coordinates $\bp_1$
and $\bp_2$) the Poincar\'e of a saddle is $-1$ (because the Jacobian has
two eigenvalues of different sign), while the Poincar\'e index of center
is $+1$ (the Jacobian has two imaginary, complex conjugate eigenvalues).
For a stable (unstable) rest-point in $p$-dimensional space the Poincar\'e index is equal to
$(-1)^{p}$ ($1$).

Recall the content of the Poincar\'e-Hopf theorem as
applied to the replicator equation \cite{hofbauer}: for regular
rest points $\pi$ of replicator dynamics (\ref{11}, \ref{12}) one has
\BEA
\label{barsuk}
{\sum}'_{\pi}{\rm Ind}[\pi] =(-1)^{n-1},
\EEA
where the sum ${\sum}'_{\pi}$ is taken over those rest points $\pi$
of (\ref{11}) which either have all their components strictly
positive $\pi_k>0$, or if some of those components are zero,
we have for each zero component $\pi_i=0$:
\BEA
\label{bosch}
\left[{\cal F}_i(\bp)-{\sum}_{\alpha=1}^n\bp_\alpha {\cal F}_\alpha(\bp)\left.\right]\right|_{\bp=\pi}<0.
\EEA
Condition (\ref{bosch}) can be rephrased by saying that if a morph $i$
is missing within the rest-point (i.e. $\pi_i=0$), then its fitness is
smaller than the mean fitness.

\comment{(ready part)
In Appendix \ref{ap1} we remind the definition of Poincar\'e indices, and
present a straightforward argument for the validity of (\ref{barsuk}),
which repeats informally the rigorous proof given in \cite{hofbauer}.
An accessible introduction to Poincar\'e indices is given in
\cite{arnold}.}

\comment{(ready part)
Let us check (\ref{barsuk}) for $n=2$ replicator equation (\ref{18})
with time-independent $A$ and $B$. The rest points of this equation can
be classified as follows: {\it (1)} one interior stable rest point (${\rm
Ind}=-1$) and two unstable rest points $p_1=1$ and $p_1=0$; both of them
have to excluded from (\ref{barsuk}) due to (\ref{bosch}), hence
(\ref{barsuk}) holds. {\it (2)} One interior unstable rest point (${\rm
Ind}=1$) and two stable rest points $p_1=1$ and $p_1=0$ with indices
${\rm Ind}=-1$. Eq.~(\ref{barsuk}) is valid. {\it (3)} No interior rest-point,
while of the boundary rest points in unstable, and another boumdary
rest-point is stable. Only the stable one with ${\rm Ind}=-1$
contributes into (\ref{barsuk}).
}

In addition to (\ref{barsuk}) we recall that the non-linear term
$b_{klm}$ and $c_{klmn}$ in the effective replicator equation (\ref{11},
\ref{12}) do not influence on the stability of the vertex rest points,
where all $\bp_k$'s besides one nullify; see our discussion in section
\ref{barra}. 

Eq.~(\ref{barsuk}) shows that the conclusion of section \ref{two} on the
simultaneous emergence of two new rest-points (one stable and another
unstable) is a general feature of (\ref{11}, \ref{12}). Indeed, for even
values of $n$, stable rest-points have ${\rm Ind}=-1$ (since the number
of independent variables is odd, and each one brings factor $-1$ into
the index). Hence according to (\ref{barsuk}), such a stable rest-point
has to emerge simultaneously with another rest-point that has at least
one unstable direction. Likewise, for odd values of $n$, where stable
rest-points have ${\rm Ind}=1$. 

Consider the three-morph situation with independent probabilities
$\bp_1$ and $\bp_2$. If the non-linear terms (\ref{14}) create a stable
rest point in the interior (polymorphism), then the sum of Poincar\'e
indices increases by $1$. So for (\ref{barsuk}) to hold, a saddle (with
Poincar\'e index $-1$) rest-point has to be created as well. Taken
together with the fact that the stability of the vertices is not
altered, this then implies that two directions of the saddle has to
joint together and form a closed curve that would separate the
attraction basin of the newly created stable rest point. This was the
scenario we saw in section \ref{three} \footnote{Due to the zero-sum
feature (\ref{250}) of the example considered in section \ref{three},
the stable rest point was only neutrally stable. However, if condition
(\ref{250}) is (slightly) relaxed one can obtain also asymptotic
stability of the polymorphic rest-point created by non-linear terms
$b_{klm}$ and $c_{klmn}$.}. 

Can the non-linear terms $b_{klm}$ and $c_{klmn}$ 
turn the stable polymorphic stable rest point|which exists
already without them|into an unstable rest-point? This
is not excluded by (\ref{barsuk}), because both rest-points
have ${\rm Ind}=1$, but it is incompatible with the
fact that the stability of the vertices is not altered by the non-linear
terms. It is though not excluded that the non-linear terms would lead to
changing of the old stable node into a unstable node surrounded by a
stable limit cycle.  Although we did not find such scenarios of the
polymorphism emergence, from the conceptual viewpoint such a scenario
will not change our basic qualitative conclusion that the poymorphism
which exists without non-linear terms is (generically) not eliminated by
them.  The same argument shows that an unstable rest point cannot
turn into a stable one without generating new rest points.

\comment{
Naturally, two of the main messages above|that the creation of an
interior stable rest point should be accompanied by creating an unstable
one, and that the stability of vertices is not altered|remain true for a
larger ($n\geq 4$) number of morphs. We shall however not dwell on this
case, because the qualitative analysis in larger-dimensional spaces is
complicated. }

\section{Summary and open problems}
\label{summa}

We discussed a mechanism of polymorphism that exists within
continuous-time population dynamics due to a rapidly-changing
(fine-grained) time-periodic environment. The characteristic time of
environmental oscillations is larger than the the time over which the
fractions of the sub-populations (morphs) change systematically (i.e. in
average) \cite{levins,clark}. Various mechanisms of polymorpshim were at
the focus of population biology for decades
\cite{levins,stein,bull,grant,kassen_minireview,wilke,dempster,janavar,svi,nagylaki_book}.
Such polymorphism scenarios as frequency-dependent selection and
heterozygote advantage are at the basis of the current population
thinking. Polymorphism induced by inhomogeneous environment also
attracted muchattention\cite{levins,stein,bull,grant,kassen_minireview,wilke,dempster,janavar,svi,nagylaki_book}.
So far this attention focussed on the slow (coarse-grained) environment,
because early arguments implied that non-trivial polymorphism scenarios
are absent for a rapidly-changing environment
\cite{levins,clark,strobeck}.  Also the earlier studies of replicator
dynamics in a time-dependent environment concentrated on those
particular cases, where non-trivial scenarios of polymorphism are indeed
absent \cite{nagylaki,kimura,li}. 

Starting from the replicator equation in a fast time-periodic
environment we got an effective replicator equation for the
time-averaged fractions of morphs.  The main difference with the
ordinary replicator equation in the time-independent environment|which
for symmetric pay-offs reduces to the Fisher equations of mathematical
genetics \cite{svi,nagylaki_book}|is that the fitness contains
additional non-linear terms that appear due to the morphs tracking the
environmental changes (the linear terms in the fitness correspond to the
time-averaged environment).  The presence of such tracking in a
rapidly-changing (fine-grained) environment is established
observationally \cite{drent}.  The form of non-linear terms allows to
draw general conclusions on the stability and the effective fitness. 

The non-linear terms can create a stable polymorphic state. But they
generically do not destroy polymorphism that exists without the presence
of these terms (i.e., for the averaged environment). Our results are
worked out for three pertinent cases: genetic selection
between two morphs (alleles), prisoner's dilemma game (where
polymorphism implies a resolution of the prisoner's dilemma due to a
time-varying environment), and the rock-paper-scissors game between
three cyclically dominating morphs. 

For the symmetric-pay-off situation [genetic selection] the fitness of
the overall population can decrease due to the polymorphism. Once the
existing polymorphism is generically not modified by the
environment-induced terms, the adaptive value of the polymorphism is in
increasing the structural stability of the population under influence of
a time-dependent environment. 

Several open problems are suggested by this research.

{
-- One should clarify to which extent the uncovered mechanism of
polymorphism is relevant for the phenomenon of sympatric speciation,
where by contrast to the allopatric scenario the speciation is induced
inside a single population \cite{doeb}. (Within the allopatric scenario
sub-populations are first isolated geographically and only then
speciate). Interestingly, the decrease of the mean fitness was argued to
be a prerequisite for the sympatric speciation \cite{doeb}. Thus, our
results hint at a sympatric speciation scenario due to a
rapidly-changing, time-periodic environment.

-- How to extend the present approach to multi-locus genetic selection,
where polymorpshim related to a slow environment was studied recently in
\cite{ppnass}? 

-- It will be interesting to develop the present approach for 
Eigen-Schuster model and its ramifications living on dynamic landscapes; see
\cite{wilke} for a review. 

-- The basic Lotka-Volterra equations of ecological dynamics can be
recast in the form (\ref{z2}) and studied as replicators
\cite{hofbauer}. Note however that despite of this mathematical
equivalence the ecological (and biological) content of Lotka-Volterra
equations differs from that of replicators. This especially concerns the
time-scale separation issues (i.e. the division between slow and fast),
because Lotka-Volterra equations have additional time-scales
\cite{svirizhev_logofet}. Though the theory developed above does not
apply directly to Lotka-Volterra equations, developing such applications
is a pertinent avenue of future research. 

-- The present approach was developed assuming that all time-dependent
selection coefficients have the same, well-defined period. This is 
restrictive an assumption, and we expect a richer dynamics upon relaxing
it. 

}

\subsection*{Acknowledgments}

\comment{
We thank M. Broom for useful remarks. K. Petrosyan K.-t.  Leung are
acknowledged for critical reading of an initial version of this work. We
were supported by Volkswagenstiftung, grants NSC 96-2911-M 001-003-MY3
\& AS-95-TP-A07, and National Center for Theoretical Sciences in Taiwan. }

AEA and SGB were supported by SCS of Armenia, grants No. 18RF-002 and No. 18T-1C090.
CKH was supported by Grant MOST 107-2112-M-259-006.

{

\appendix

\section{Replicator equation under condition (\ref{mashad}) with second-order terms}
\label{mansur}

The purpose of this Appendix is to extend the asymptotic method of
section \ref{deriv} to the order ${\cal O}(\omega^{-2})$, and thereby to
understand implications of condition (\ref{mashad}). Recall that under 
(\ref{mashad}), the non-linear terms (\ref{14}) in the fitness are absent. 
For simplicity we shall work with $n=2$ (two morphs). 

Our first conclusion is that under condition (\ref{mashad}) the terms
${\cal O}(\omega^{-2})$ that modify the replicator equation are 
perturbative|i.e. they have to be smaller than the terms given by the
time-averages of pay-offs (selection coefficients). Our second
conclusion (closely related to the first one) is that these terms are
not useful for polymorphism. 

Consider 
\BEA
\label{dottir}
\dot{p}=p(1-p)[A(\tau)-B(\tau)p],~~ \tau=\omega t, ~ 0\leq p(t)\leq 1,~
\EEA
where $A(\tau)$ and $B(\tau)$ are rapidly-changing functions of time due
to a large $\omega$.  The time-averages, as well as the constant and
oscillating parts of $A$ and $B$ are defined according to
(\ref{corrida}--\ref{darder}). 

Following the method of section \ref{ee}, we look for the solution  
$p(t)$ of (\ref{dottir}) as [cf.~(\ref{8})]
\begin{gather}
\label{zoro}
p(t)=\bp(t)+\frac{\epsilon_1[\bp(t),\tau]}{\omega}+\frac{\epsilon_2[\bp(t),\dot{\bp}(t),\tau]}{\omega^2}+{\cal O}(\frac{1}{\omega^3}),
\end{gather}
where the time-averages hold [cf.~(\ref{8.1})]
\BEA
\bar{\epsilon}_1=\bar{\epsilon}_2=0.
\EEA

Putting (\ref{zoro}) into (\ref{dottir}) we get
\BEA
\label{she1}
&&\dot{\bp}(t)+\frac{\dot{\bp}(t) \epsilon_1'}{\omega}+\frac{\dot{\bp}(t) \epsilon_2'}{\omega^2}+\partial_{\tau}\epsilon_1
+\frac{1}{\omega}\partial_{\tau}\epsilon_2\\
\label{she2}
&&=f[\bp,A(\tau),B(\tau)]+\frac{\epsilon_1}{\omega} f'[\bp, A(\tau), B(\tau)]\\
\label{she3}
&&+\frac{\epsilon_2}{\omega^2} f'[\bp, A(\tau), B(\tau)]
+\frac{\epsilon^2_1}{2\omega^2} f''[\bp, A(\tau), B(\tau)],
\EEA
where we denoted 
\BEA
\label{raa}
f(p, A, B)\equiv p(1-p)[A-Bp], 
\EEA
and $\epsilon'$ and $f'$ (resp. $\epsilon''$ and $f''$) 
means the first (resp. second) derivative over the first argument. 

On the fast times we collect from (\ref{she1}) the orders of ${\cal O}(1)$ and ${\cal O}(1/\omega)$ respectively
[cf.~(\ref{z8z}, \ref{z8})]:
\BEA
\label{krot1}
&& \partial_\tau\epsilon_1=f[\bp,\widetilde{A}, \widetilde{B}],\\
&& \partial_\tau\epsilon_2=-\dot{\bp}\epsilon_1'+\epsilon_1 f[\bp,\widetilde{A}, \widetilde{B}].
\label{krot2}
\EEA
Before solving (\ref{krot1}, \ref{krot2}), let us assume [cf.~(\ref{mashad})]:
\BEA
A(\tau)=\bar{A}+a h(\tau), \qquad A(\tau)=\bar{B}+b h(\tau),
\EEA 
where $h(\tau)=h(\tau+2\pi)$ is a periodic function of $\tau$, and where $\bar{A}$, $\bar{B}$, $a$ and $b$
are constants. 

Then (\ref{krot1}, \ref{krot2}) are solved as
\BEA
\label{guf1}
\epsilon_1&=&f(\bp,a,b)\hat{h}(\tau), \\
\epsilon_2&=&[\, f(\bp,a,b)f'(\bp,\bar{A},\bar{B})-\dot{\bp} f'(\bp,a,b)   \,]\,\hat{\hat{h}}(\tau)\nonumber\\
&&+\frac{1}{2}\,f(\bp,a,b)f'(\bp,a,b)\left(\hat{h}^2(\tau) - \overline{\,\hat{h}^2\,}\, \right),
\label{guf2}
\EEA
where ${\hat{h}}(\tau)$ and $\hat{\hat{h}}(\tau)$ are the first and second primitive of $h(t)$. Both hold the zero average condition
$\overline{\, {\hat{h}}\,}= \overline{\,\hat{\hat{h}}\,}=0$.

Putting (\ref{guf1}) and (\ref{guf2}) into (\ref{she2}, \ref{she3}) and taking the time-average, we get
\begin{gather}
\left(1-\chi\, [f'(\bp,a,b)\,]^2  \,\right) \dot{\bp}=f(\bp,\bar{A},\bar{B})+\chi f(\bp,a,b)\times\nonumber\\
\label{baru}
\left[\frac{1}{2}f''(\bp,\bar{A},\bar{B})f(\bp,a,b)  
-f'(\bp,\bar{A},\bar{B})f'(\bp,a,b) \,\right],\\
\chi\equiv \overline{\, \hat{h}\,\hat{h} \,}/\omega^2.
\label{baruch}
\end{gather}
In contrast to (\ref{11}), where non-linear fitness terms need not be
small, Eq.~(\ref{baru}) shows that the non-linear fitness terms|those
proportional to $\chi$|are smaller than the average fitness term
$f(\bp,\bar{A},\bar{B})$. This is seen most clearly, when we put
$\bar{A}=\bar{B}=0$ and notice that the right-hand-side of (\ref{baruch})
disappears: 
\BEA
\label{ushi}
\left(1-\chi\, [f'(\bp,a,b)\,]^2  \,\right) \dot{\bp}=\dot{\bp}=0.
\EEA
One can still ask whether the terms $\propto\chi$ can lead to polymorphism 
in the boundary situation, e.g. given $\bar{A}$ a positive but small (i.e. when $\bp=0$
is marginally stable), can those terms change the stability of these boundary rest-point. 
The answer is negative.
}


\begin{thebibliography}{99}

\bibitem{levins} R. Levins, {\it Evolution in Changing
Environments} (Princeton University Press, 1968).

\bibitem{stein} 
P.W. Hedrick, M.E. Ginevan and E.P. Ewing, {\it Genetic Polymorphism in
Heterogeneous Environments}, Ann. Rev. Ecol. Syst. {\bf 7}, 1 (1976). 

P.W. Hedrick, {\it ibid}, {\it Genetic Polymorphism in Heterogeneous
Environments: A Decade Later}, {\bf 17}, 535 (1986); {\it ibid}, Genetic
Polymorphism in Heterogeneous Environments: The Age of Genomics, {\bf
37}, 67-93 (2007). 

\bibitem{bull} L.A. Meyers and J.J. Bull, 
{\it Fighting change with change: adaptive variation in an uncertain world},
Tr. Ecol. Evol {\bf 17}, 551-557 (2002).

\bibitem{grant}V. Grant, {\it Organismic Evolution} (Freeman, SF, 1977).

\bibitem{kassen_minireview}R. Kassen, {\it The experimental evolution of
specialists, generalists, and the maintenance of diversity}, J. Evol.
Biol. {\bf 15}, 173 (2002). 

\bibitem{wilke} C.O. Wilke, C. Ronnewinkel, and T. Martinetz, {\it
Dynamic Fitness Landscapes in Molecular Evolution}, Physics Reports,
{\bf 349}, 395-446 (2001). 

\comment{We study self-replicating molecules under externally varying
conditions. Changing conditions such as temperature variations and/or
alterations in the environment's resource composition lead to both
non-constant replication and decay rates of the molecules. In general,
therefore, molecular evolution takes place in a fluctuating rather than
a static fitness landscape. We incorporate dynamic replication and decay
rates into the standard quasispecies theory of molecular evolution, and
show that for periodic time-dependencies, a system of evolving molecules
enters a limit cycle for $t\to\infty$. For fast periodic changes, we show that
molecules adapt to the time-averaged fitness landscape, whereas for slow
changes they track the variations in the landscape arbitrarily closely.
We derive a general approximation method that allows us to calculate the
attractor of time-periodic landscapes, and demonstrate using several
examples that the results of the approximation and the limiting cases of
very slow and very fast changes are in perfect agreement. We also
discuss landscapes with arbitrary time dependencies, and show that very
fast changes again lead to a system that adapts to the time-averaged
landscape. Finally, we analyze the dynamics of a finite population of
molecules in a dynamic landscape, and discuss its relation to the
infinite population limit.}

\bibitem{dempster} E. Dempster, {\it Maintenance of genetic heterogeneity}, Cold Spring Harbor Symp. Quant. Biol. {\bf 20}, 25 (1955).


\bibitem{janavar} 
J.B.S. Haldane and S.D. Jayakar, {\it Polymorphism due to selection of varying direction}, J. Genet. {\bf 58}, 237 (1963).

J.L. Cornette, {\it Deterministic genetic models in varying environments}, J. Math. Biol. {\bf 12}, 173-186 (1981).


\bibitem{svi}Yu.M. Svirezhev and V.P. Passekov, {\it Findamentals of Mathematical
Genetics} (Dordrecht, Kluwer, 1990).

\bibitem{nagylaki_book}T. Nagylaki, {\it Introduction to Theoretical Population Genetics}
(Springer-Verlag, Berlin, 1992).

\bibitem{nagylaki} T. Nagylaki, {\it Polymorphisms in cyclically-varying
environments}, Heredity, {\bf 35}, 67 (1975). 


\bibitem{kimura}M. Kimura, {\it Stochastic processes and distribution of gene sequences under natural selection},
Cold Spring Harbor Symp. Quant. Biol., 20, 33 (1955).


\bibitem{li}W.-H. Li, {\it Retention of cryptic genes in microbial
populations}, Mol. Biol. Evol., {\bf 1}, 213-219 (1984). 

\bibitem{gill} J.H. Gillespie, {\it The Causes of Molecular Evolution}
(Oxford Univ. Press, Oxford, 1991).

\bibitem{clark} {\it Lecture Notes on Biomathematics:
Adaptation in Stochastic Environments}, ed.
J. Yoshimura and C.W. Clark (Springer-Verlag, Berlin, 1991).

\bibitem{strobeck}C. Strobeck,{\it Selection in a fine-grained
environment}, Am. Nat. {\bf 109}, 419-425 (1975). 

\bibitem{miner_vonesh}B.G. Miner and J.R. Vonesh, {\it Effects of fine
grain environmental variability on morphological plasticity}, Ecol.
Lett. {\bf 7}, 794-801 (2004). 

N. M. Schoeppner and R. A. Relyea, {\it Phenotypic plasticity in response to fine-grained
environmental variation in predation}, Functional Ecology, {\bf 23}, 587 (2009).

\bibitem{winn}A. Winn, {\it Adaptation to fine-grained environmental
variation: an analysis of within-individual leaf variation in an annual
plant}, Evolution {\bf 50}, 1111 (1996). 

\bibitem{jasmin}J.-N. Jasmin and R. Kassen, {\it On the experimental
evolution of specialization and diversity in heterogeneous
environments}, Proc. R. Soc. B {\bf 274}, 2761 (2007). 

\bibitem{cook}L.M. Cook, {\it A two-stage model for Cepaea polymorphism}, Phil. Trans. R. Soc. B {\bf 353}, 1577 (1998).

\bibitem{zakharov}I.A. Zakharov, {\it Red and Black}, Priroda, {\bf 5}, 46 (1992). (In Russian)

\bibitem{lady} M.E.N.Majerus and P.W.E. Kearns, {\it Ladybirds} (Slough, London, 1989).


\bibitem{aversion}
L.A. Real, {\it Fitness uncertainty and the role of diversification in
evolution and behavior}, Am. Nat. {\bf 115}, 623-638 (1980).


H. A. Orr, Evolution, {\it Absolute fitness, relative fitness and utility}, {\bf 61}, 2997-3000 (2007).

\bibitem{ford}E.B. Ford, {\it Ecological genetics} (Chapman and Hall, London, 1975) 

\bibitem{drent}T. Piersma and J. Drent, {\it Phenotypic flexibility and the evolution of organismal design},
Trends in Ecology and Evolution, {\bf 18}, 228 (2003).

\bibitem{prl} A. E. Allahverdyan and C.-K. Hu, Phys. Rev. Lett. {\bf 102}, 058102 (2009).

\bibitem{hofbauer}
J. Hofbauer and K. Sigmund, {\it Evolutionary Games and Population
Dynamics} (Cambridge Univ. Press, 1998).

\bibitem{zeeman} E.C. Zeeman, {\it Population dynamics from game
theory}, in Z. Nitecki and C. Robinson (editors), {\it Global Theory of
Dynamical Systems} (Lecture Notes in Mathematics, {\bf 819}, Springer,
Berlin).  DOI: https://doi.org/10.1007/BFb0087009

\bibitem{levins79}
R. Levins, {\it Coexistence in a Variable Environment}, The American Naturalist, {\bf 114}, 765-783 (1979).

\bibitem{namara}J.M. McNamara, {\it Towards a richer evolutionary game theory},
J. R. Soc. Interface, {\bf 10}, 20130544 (2013).


\bibitem{fox}J. W. Fox, {\it The intermediate disturbance hypothesis should be abandoned},
Trends in Ecology \& Evolution, {\bf 28}, 86-92 (2013).

{
\bibitem{sheil}D. Sheil and D.F. Burslem, {\it Defining and defending
Connell's intermediate disturbance hypothesis: a response to Fox},
Trends in Ecology \& Evolution, {\bf 28}, 571-572 (2013). 
}


\bibitem{thomas} B. Thomas, {\it Evolutionary stability: states and strategies}, Theoretical Population Biology {\bf 26}, 49 (1984).

\comment{C. T. Bergstrom and P. Godfrey-Smith, Biology and Philosophy, {\bf 13}, 205 (1998). }



\bibitem{ll}
L. D. Landau and E. M. Lifshitz, {\it Mechanics} (Pergamon
Press, Oxford, 1976).

\bibitem{kapitza} P. L. Kapitza, {\it Dynamic stability of a pendulum
with an oscillating point of suspension}, Zh. Eksp. Teor. Fiz. 21, 588
(1951). 

\bibitem{davidson} A. Ridinger and N. Davidson, {\it Particle motion in
rapidly oscillating potentials: The role of the potential's initial
phase}, Phys. Rev. A {\bf 76}, 013421 (2007). 

\bibitem{slips} F. Haake and M. Lewenstein, {\it Adiabatic drag and
initial slip in random processes}, Phys. Rev. A {\bf 28}, 3606 (1983). 

S. M. Cox and A. J. Roberts, {\it Initial conditions for models of dynamical systems},
Physica D {\bf 85}, 126 (1995).

\comment{The long-time behaviour of many dynamical systems may be
effectively predicted by a low-dimensional model that describes the
evolution of a reduced set of variables. We consider the question of how
to equip such a low-dimensional model with appropriate initial
conditions, so that it faithfully reproduces the long-term behaviour of
the orginal high-dimensional dynamical system. Our method involves
putting the dynamical system into normal form, which not only generates
the low-dimensional model, but also provides the correct initital
conditions for the model. We illustrate the method with several
examples.}


\bibitem{broom_vick}M. Broom, C. Cannings and G.T. Vickers, Bull. Math.
Biol., {\it Multi-player matrix games}, {\bf 59}, 931-952 (1997). 

L.A. Bach, T. Helvik and F.B. Christiansen, {\it The evolution of
n-player cooperation—threshold games and ESS bifurcations}, J. Theor.
Biol. {\bf 238}, 426-434 (2006). 

\bibitem{shubik}M. Shubik, {\it Game Theory, Behavior and the Paradox of
the Prisoner's Dilemma: Three Solutions}, Journal of Conflict
Resolution, {\bf 14}, 181 (1970).

\bibitem{peterson}{\it The Prisoner's Dilemma}, M. Peterson (ed.)
(Cambridge University Press, Cambridge, 2015).


\bibitem{sandler} D. G. Arce and T. Sandler, {\it The Dilemma of the
Prisoners' Dilemmas}, KYKLOS {\bf 58}, 3-24 (2005). 

\bibitem{dyson} W. H. Press and F. J. Dyson, {\it Iterated Prisoner's
Dilemma contains strategies that dominate any evolutionary opponent},
Proc. Natl. Acad. Sci. U.S.A, {\bf 109}, 10409 (2012). 

\bibitem{15saakian} T.Yakushkina, D. B. Saakian, A. Bratus, and C.-K.
Hu, {\it Evolutionary games with randomly changing payoff matrices}, J.
Phys. Soc. Japan {\bf 84}, 064802 (2015). 

\bibitem{benica} E. Benincà, B. Ballantine, S.P. Ellner, and J. Huisman,
{\it Species fluctuations sustained by a cyclic succession at the edge
of chaos}, Proc. Natl. Acad. Sci. U.S.A, {\bf 112}, 6389-6394 (2015). 

\bibitem{weitz}J.S. Weitz {\it et al.}, {\it An oscillating tragedy of
the commons in replicator dynamics with game-environment feedback},
Proc. Natl. Acad. Sci. U.S.A., {\bf 113}, E7518-E7525 (2016). 



\comment{

J. Felsenstein, {\it The theoretical population genetics
of variable selection and migration}, Ann. Rev. Genet. {\bf 10} 253
(1976). 



\bibitem{paul}
W. Paul, Rev. Mod. Phys. {\bf 62}, 531 (1990).

\bibitem{rahav}
S. Rahav, I. Gilary, and S. Fishman, Phys. Rev. A {\bf 68}, 013820 (2003).


S.C. Stearns, {\it Daniel Bernoulli (1738): evolution and economics under risk},
J. Biosci. {\bf 25}, 221-228 (2000).




\bibitem{unto}E. Sober and D. S. Wilson, {\it Unto Others: The Evolution
and Psychology of Unselfish Behavior} (Harvard University Press,
Cambridge, MA, 1998).


\bibitem{rand} R.E. Ruelas, D. G. Rand, and R. H. Rand, {\it Parametric
excitation and evolutionary dynamics}, Journal of Applied Mechanics {\bf
80}, 050903 (2013). 

\bibitem{broom}
M. Broom, Comp. Rend. Biol. {\bf 328}, 403 (2005).


\bibitem{traulsen} W. Huang {\it et al.}, {\it Emergence of stable polymorphisms driven by evolutionary games between mutants},
Nature Communications {\bf 3}, 919 (2012).

\bibitem{david}D. B. Saakian, A. Bratus, and  C.-K. Hu, {\it Biological evolution model with conditional mutation rates},
Physica A, {\bf 474}, 32-38 (2017).


\bibitem{smith}
J. Maynard Smith, {\it Evolution and the Theory of Games} (Cambridge
University Press, 1982).

D.W. Stephens and J.R. Krebs, {\it Foraging Theory}
(Princeton University Press, 1986).


\bibitem{brandon}R.N. Brandon, {\it Adaptation and environment } (Princeton University Press, 1990).
}



\bibitem{tanimoto} J. Tanimoto, {\it Dilemma solving by the coevolution
of networks and strategy in a $2\times 2$ game}, Phys. Rev. E, {\bf 76},
021126 (2007). 

\bibitem{helbing}
D. Helbing and S. Lozano, {\it Phase transitions to cooperation in the 
prisoner's dilemma}, Phys. Rev. E, {\bf 81}, 057102 (2010).

\comment{
Game theory formalizes certain interactions between physical particles
or between living beings in biology, sociology, and economics and
quantifies the outcomes by payoffs. The prisoner's dilemma (PD)
describes situations in which it is profitable if everybody cooperates
rather than defects (free rides or cheats), but as cooperation is risky
and defection is tempting, the expected outcome is defection.
Nevertheless, some biological and social mechanisms can support
cooperation by effectively transforming the payoffs. Here, we study the
related phase transitions, which can be of first order (discontinuous)
or of second order (continuous), implying a variety of different routes
to cooperation. After classifying the transitions into cases of
equilibrium displacement, equilibrium selection, and equilibrium
creation, we show that a transition to cooperation may take place even
if the stationary states and the eigenvalues of the replicator equation
for the PD stay unchanged. Our example is based on adaptive group
pressure, which makes the payoffs dependent on the endogenous dynamics
in the population. The resulting bistability can invert the expected
outcome in favor of cooperation. }

\bibitem{devlin}
S. Devlin and T. Treloar, {\it Network-based criterion for the success 
of cooperation in an evolutionary prisoner's dilemma}, Phys. Rev. E, 
{\bf 86}, 026113 (2012).

\comment{
We consider an evolutionary prisoner's dilemma on a random network. We
introduce a simple quantitative network-based parameter and show that it
effectively predicts the success of cooperation in simulations on the
network. The criterion is shown to be accurate on a variety of networks
with degree distributions ranging from regular to Poisson to scale free.
The parameter allows for comparisons of random networks regardless of
their underlying topology. Finally, we draw analogies between the
criterion for the success of cooperation introduced here and existing
criteria in other contexts. }



\bibitem{abraham}M. Frean and E. R. Abraham, {\it Rock-scissors-paper
and the survival of the weakest}, Proc. R. Soc. Lond. B {\bf 268},
1323-1327 (2001). 

\bibitem{sinervo} B. Sinervo and C. M. Lively, {\it The
rock-paper-scissors game and the evolution of alternative male
strategies}, Nature, {\bf 340}, 240 (1996). 

\bibitem{buss} L. W. Buss, {\it Competitive intransitivity and
size-frequency distributions of interacting populations}, Proc. Natl.
Acad. Sci. U.S.A, {\bf 77}, 5355 (1980). 



\bibitem{kerr} B. C. Kirkup and M. A. Riley, {\it Antibiotic-mediated
antagonism leads to a bacterial game of rock-paper-scissors in vivo},
Nature {\bf 428}, 412 (2004). 

B. Kerr, M. A. Riley, M. W. Feldman, and B. J. M. Bohannan, {\it Local
dispersal promotes biodiversity in a real-life game of
rock-paper-scissors}, Nature, {\bf 418}, 171 (2002). 


\bibitem{schuster}G. Neumann and S. Schuster, {\it Continuous model for
the rock-scissors-paper game between bacteriocin producing bacteria}, J.
Math. Biol. {\bf 54}, 815 (2007). 






\bibitem{akin} E. Akin and V. Losert, {\it Evolutionary dynamics of zero-sum games},
J. Math. Biology {\bf 20}, 231 (1984).





\bibitem{doeb}M. Doebeli and U. Dieckmann, {\it Evolutionary branching
and sympatric speciation caused by different types of ecological
interactions}, Am. Nat. {\bf 156}, S77 (2000); J. Evol. Biol. {\it
Adaptive dynamics as a mathematical tool for studying the ecology of
speciation processes}, {\bf 18}, 1194-1200 (2005). 



\bibitem{ppnass} M. J. Wittmann, A. O. Bergland, M. W. Feldman, P. S. Schmidt, and
D. A. Petrov, {\it Seasonally fluctuating selection can maintain polymorphism at many loci via segregation lift}, 
Proc. Natl. Acad. Sci. U.S.A, {\bf 114}, E9932–E9941 (2017).

\comment{ 
Most natural populations are affected by seasonal changes in
temperature, rainfall, or resource availability. Seasonally fluctuating
selection could potentially make a large contribution to maintaining
genetic polymorphism in populations. However, previous theory suggests
that the conditions for multilocus polymorphism are restrictive. Here,
we explore a more general class of models with multilocus seasonally
fluctuating selection in diploids.  In these models, the multilocus
genotype is mapped to fitness in two steps. The first mapping is
additive across loci and accounts for the relative contributions of
heterozygous and homozygous loci—that is, dominance. The second step
uses a nonlinear fitness function to account for the strength of
selection and epistasis. Using mathematical analysis and
individual-based simulations, we show that stable polymorphism at many
loci is possible if currently favored alleles are sufficiently dominant.
This general mechanism, which we call "segregation lift," requires
seasonal changes in dominance, a phenomenon that may arise naturally in
situations with antagonistic pleiotropy and seasonal changes in the
relative importance of traits for fitness. Segregation lift works best
under diminishing-returns epistasis, is not affected by problems of
genetic load, and is robust to differences in parameters across loci and
seasons. Under segregation lift, loci can exhibit conspicuous seasonal
allele-frequency fluctuations, but often fluctuations may be small and
hard to detect. An important direction for future work is to formally
test for segregation lift in empirical data and to quantify its
contribution to maintaining genetic variation in natural populations.

Significance 
A key question in evolutionary biology is: What maintains
the abundant genetic variation observed in natural populations? Many
organisms experience some seasonality in their habitats, and, if they
have multiple generations per year, seasonally fluctuating selection is
a potentially powerful mechanism to maintain polymorphism. However,
previous research has argued that this occurs rarely. Inspired by recent
empirical findings, we reevaluate the potential of seasonally
fluctuating selection to simultaneously maintain polymorphism at many
loci in the genome. We obtain a more general condition for the
maintenance of multilocus polymorphism by seasonally fluctuating
selection. This condition may plausibly be satisfied for many species
and does not suffer from problems of previous models. 

}



\bibitem{svirizhev_logofet} Yu.M. Svirizhev and D.O. Logofet, {\it
Stability of biological communities} (Nauka, Moscow, 1978) (In Russian). 



\end{thebibliography}
\end{document}